# Electronic Self-Organization in the β-Pyrochlore Oxide CsW$_2$O$_6$

Yoshihiko Okamoto*

*Institute for Solid State Physics, University of Tokyo, Kashiwa 277-8581, Japan*

In this review, I present the electronic properties of the β-pyrochlore oxide CsW$_2$O$_6$ and other related materials. At 215 K, CsW$_2$O$_6$ exhibits an electronic phase transition to a nonmagnetic insulating state, which exhibits a complex self-organization of 5$d$ electrons. In this phase transition, various factors, such as geometrical frustration of the pyrochlore structure, moderately strong electron correlation, Jahn–Teller-like distortion, phase transition to the insulating state, and formation of chemical bonds in solids, which are powerful driving forces for achieving a wide variety of electronic properties in transition metal oxides, play important roles. In CsW$_2$O$_6$, the interplay of these factors has led to the emergence of an electronic phase transition that preserves the cubic symmetry, three-dimensional nesting of Fermi surfaces, a possible charge order with a fractional valence satisfying the Anderson condition, and an equilateral-triangular trimer formation by a three-centered-two-electron bond. In addition to the introduction of these unique features of CsW$_2$O$_6$, each of them has been compared to those of other materials to provide an overview of the electronic properties of a wide variety of related materials, which can contribute to a complete understanding of the vast and infinite electronic phenomena in transition metal oxides.

**1. Introduction**

Pyrochlore was originally the name of an equiaxed mineral with the chemical composition (Na,Ca)$_2$Nb$_2$O$_6$(OH,F) [1]. In condensed matter physics, the word pyrochlore has several different meanings. First, it is the name of a materials group that has the same crystal structure as the pyrochlore minerals and the chemical composition A$_2$B$_2$X$_6$X′, where A and B are metallic elements and X and X′ are negative elements, such as oxygen and fluorine. In this review, this crystal structure and the oxides with this crystal structure are called α-pyrochlore structure and α-pyrochlore oxides, respectively [2]. As shown in Fig. 1(a), in the α-pyrochlore oxide, the B atom is coordinated with six oxygen atoms, whereas the A atom is coordinated with eight oxygen atoms (specifically, 2 + 6 coordination of two X′ sites and six X sites) [3]. The BO$_6$ octahedra are three-dimensionally connected by sharing their corners, resulting in a three-dimensional array of corner-shared tetrahedra of B atoms [Fig. 1(c)]. This B sublattice is often called the pyrochlore lattice or pyrochlore structure; however in this review, it is referred to as the pyrochlore structure to distinguish it from the α- and subsequently described β-pyrochlore structures. In the α-pyrochlore structure, the A atoms also form a pyrochlore structure.

Unlike the α-pyrochlore oxide A$_2$B$_2$O$_7$, another pyrochlore oxide family exists with the chemical composition of AB$_2$O$_6$. Although this family is sometimes called modified pyrochlore, it is referred to as β-pyrochlore in this review [2,4,5]. The crystal structure of β-pyrochlore is closely related to that of the α-pyrochlore. As shown in Fig. 1(b), B atoms in β-pyrochlore are octahedrally coordinated with oxygen atoms and form a pyrochlore structure, as shown in Fig. 1(c), which is similar to the α-pyrochlore. In contrast, A atoms in β-pyrochlore occupy the X′ site of the α-pyrochlore structure, instead of the A site, and are coordinated with 18 oxygen atoms, forming a diamond structure. If not distorted, both α- and β-pyrochlore structures have a cubic $Fd\bar{3}m$ space group. In this review, α- and β-pyrochlore oxides are collectively referred to as pyrochlore oxides.

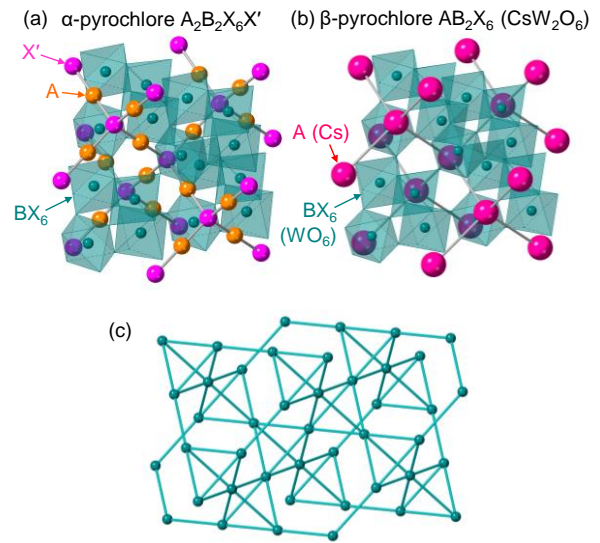

Figure 1. (a) Crystal structure of α-pyrochlore A$_2$B$_2$X$_6$X′. (b) Crystal structure of β-pyrochlore AB$_2$X$_6$ and β-pyrochlore oxide CsW$_2$O$_6$. In (a) and (b), X atoms are located at the corners of BX$_6$ octahedra. (c) Pyrochlore structure.

Pyrochlore oxides are major research targets in solid state physics because of their superior electronic properties [3,6]. They exhibit varied electronic properties, which results in a wide variety of electronic phenomena, such as spin ice [7,8], all-in-all-out magnetic order as an extended octupolar order [9–12], metal-insulator transitions [13–20], superconductivity [2,4,21–23], and topological semimetals [24,25]. Almost all these phenomena, except for spin ice, in which lanthanoid 4$f$ electrons play an essential role, are attributed to the 4$d$ and 5$d$ electrons of the 4$d$ and 5$d$ transition metal atoms in the B site, respectively. In pyrochlore oxides, the B site is often occupied by 4$d$ and 5$d$ transition metal atoms rather than 3$d$ transition metal atoms, because the B atom has a high valence of 4 to 6 [3]. However, the electron correlation in pyrochlore oxides is moderately strong for the 4$d$ and 5$d$ electron systems because of the small orbital overlaps between transition metal 4$d$/5$d$ orbitals and oxygen 2$p$ orbitals due to the bent B–O–B bond. The interplay between this moderately strong electron correlation, strong spin–orbit coupling, and geometry of the pyrochlore structure is a source of the various electronic phenomena emerging in pyrochlore oxides.

In this review, I focus on the β-pyrochlore oxide $CsW_2O_6$, which was first synthesized by Cava *et al.* in the form of a powder sample [26]. Based on the powder X-ray diffraction (XRD) data obtained at room temperature, $CsW_2O_6$ was reported to have the β-pyrochlore structure with the cubic $Fd\bar{3}m$ symmetry. As shown in Fig. 1(b), W atoms, which have an average valence of 5.5+ and 5$d^{0.5}$ electron configuration, occupy the B sites and form a pyrochlore structure. Subsequent electrical resistivity measurements using polycrystalline samples revealed a metal-insulator transition at 210 K [27]. In this study, the low-temperature insulating phase was shown to have an orthorhombic $Pnma$ space group based on synchrotron powder XRD and powder neutron diffraction data; however, this space group was suggested to be incorrect by a subsequent theoretical study [28]. Photoemission spectroscopy measurements of the synthesized thin films suggested charge disproportionation of the W atoms to the $W^{5+}$ and $W^{6+}$ states in the insulating phase [29].

Single crystals of $CsW_2O_6$ with sizes of up to several millimeters were synthesized [30,31]. Structural and physical property measurements using single crystals indicated that a phase transition preserving the cubic symmetry occurs at $T_t$ = 215 K, in which the space group changes from $Fd\bar{3}m$ to $P2_13$ and a nonmagnetic insulating state is realized below $T_t$. In this state, all the 5$d$ electrons of the W atoms are confined in the form of equilateral-triangular $W_3$ trimers because of the possible charge ordering of $W^{5.33+}$ and $W^{6+}$ and 5$d$ orbital arrangement associated with the large distortion of the $WO_6$ octahedra. This is an unprecedented self-organization of 5$d$ electrons that can be understood as the formation of a "molecule" in a solid. In this review, the studies on $CsW_2O_6$ are discussed in Chapters 2 and 3, and $CsW_2O_6$ is compared with related materials, such as pyrochlore oxides, spinels, and trimer-forming materials, in Chapter 4, which provides a bird's-eye view of the rich physics emerging in pyrochlore structures, thus highlighting the unique electronic properties of $CsW_2O_6$.

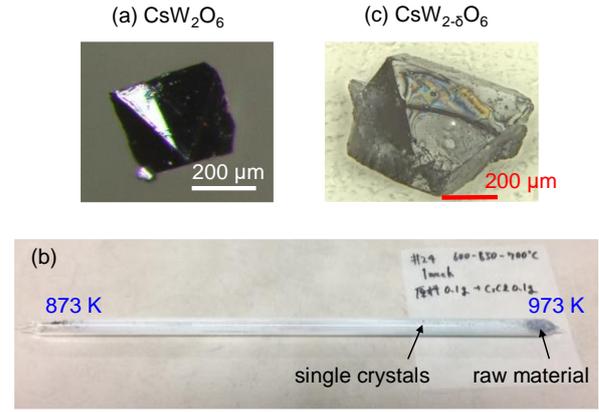

Figure 2. (a) A single crystal of $CsW_2O_6$. (b) Quartz tube after the single-crystal growth of $CsW_2O_6$. (c) A single crystal of $CsW_{2–δ}O_6$.

**2. Synthesis of $CsW_2O_6$**

Single crystals of $CsW_2O_6$ were synthesized via the crystal growth in evacuated quartz tubes under a temperature gradient. Details on the synthesis conditions are described in Refs. 30 and 31. A mixture of 3:1:3 molar ratio of $Cs_2WO_4$, $WO_3$, and $WO_2$ powders was ground, mixed, and sealed in an evacuated quartz tube containing CsCl powder. The hot and cold sides of the tube were heated to and then maintained for 96 h at 973 and 873 K, respectively; then, the furnace was cooled to room temperature. The mixture was placed on the hot side. The obtained single crystal, as shown in Fig. 2(a), has an octahedral shape with {111} planes and a maximum size of several millimeters. The molar ratios of the raw materials and reaction temperatures were based on the synthesis conditions of the powder samples, as reported in Refs. 26 and 27. A problem with this crystal growth method is that the growth conditions have not been optimized because the mechanism of crystal growth is unclear. As shown in Fig. 2(b), single crystals were grown at the middle of the quartz tube, indicating the involvement of the vapor phase in crystal growth. However, CsCl is not a commonly used transport agent for chemical vapor transport, although it is often used as a flux in the flux method. In fact, the melting point of CsCl is 918 K; therefore, CsCl is in the liquid phase at the hot side of the quartz tube, suggesting that CsCl can also play the role of a flux in this single-crystal growth.

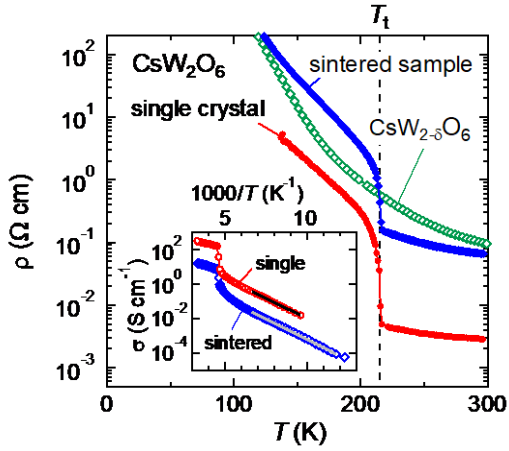
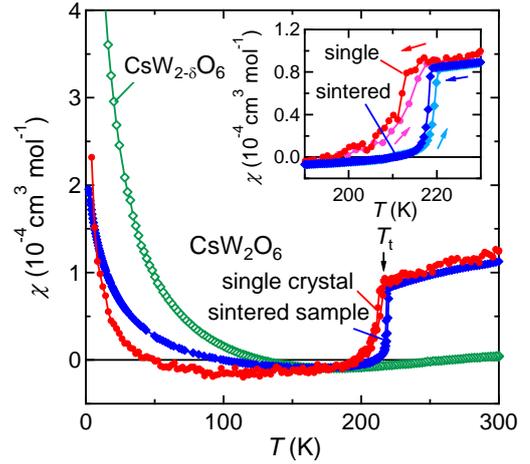

Figure 3. Temperature dependence of electrical resistivity of single crystals of $CsW_2O_6$ and $CsW_{2-\delta}O_6$ and a sintered sample of $CsW_2O_6$. Filled and open symbols indicate the data for $CsW_2O_6$ and $CsW_{2-\delta}O_6$, respectively. The broken line indicates $T_t$ = 215 K. The inset shows Arrhenius plots of electrical conductivity of single-crystalline and sintered samples of $CsW_2O_6$. Solid lines represent the results of linear fits to 104–150 and 85–150 K data for single-crystalline and sintered samples, respectively. The data were obtained from Ref. 30. © 2020 The Authors and Ref. 31. © 2020 The Physical Society of Japan.

Figure 4. Temperature dependence of magnetic susceptibility of single crystals of $CsW_2O_6$ and $CsW_{2-\delta}O_6$ and a sintered sample of $CsW_2O_6$ measured under a magnetic field of 1 T. Filled and open symbols indicate the data for $CsW_2O_6$ and $CsW_{2-\delta}O_6$, respectively. The inset shows an enlarged data around $T_t$. The data were obtained from Ref. 30. © 2020 The Authors.

Notably, attempts were made to synthesize single crystals of $CsW_2O_6$ using a conventional flux method without a temperature gradient before developing the aforementioned single-crystal growth under a temperature gradient. The same raw materials were placed in an alumina crucible and sealed in an evacuated quartz tube. The tube was heated to 923 K, slowly cooled to 873 K at a rate of −0.5 K h$^{-1}$, and then the furnace was cooled to room temperature, yielding single crystals with the β-pyrochlore structure. In this case, CsCl acted as a flux. As shown in Fig. 2(c), the obtained single crystal exhibits an octahedral shape and is larger than the crystal that is synthesized when applying a temperature gradient. However, although significantly few single crystals of $CsW_2O_6$ were observed among the small single crystals, the chemical composition of almost all single crystals was $CsW_{2-\delta}O_6$ with a significant amount of defects at the W site [30]. The value of the W defects, δ, was estimated to be 0.165 via a structural analysis using the single-crystal XRD data. In this case, the chemical composition was $CsW_{1.835}O_6$ with an average W valence of 5.99+, indicating that the W atom has no 5$d$ electron. As shown in Figs. 3 and 4, $CsW_{1.835}O_6$ does not exhibit a phase transition at $T_t$. Electrical resistivity increased with decreasing temperature below room temperature, and the magnetic susceptibility at room temperature was almost identical to that of $CsW_2O_6$ below $T_t$, indicating that $CsW_{1.835}O_6$ is a nonmagnetic insulator over the entire temperature range below room temperature.

Thus, the application of a temperature gradient suppresses the formation of $CsW_{2-\delta}O_6$ for certain reasons, which is effective in the single-crystal growth of $CsW_2O_6$, although this method is not perfect. Single crystals of $CsW_{2-\delta}O_6$ often coexist with $CsW_2O_6$ even in the case with the temperature gradient, which is indistinguishable from those of $CsW_2O_6$. However, the amount of W defect did not show continuous variation and the single crystals tend to have δ that is approximately equal to 0 or 0.165. Therefore, single crystals of $CsW_2O_6$ can be separated from those of $CsW_{2-\delta}O_6$ by measuring the magnetic susceptibility, and the single crystals confirmed to be $CsW_2O_6$ were used for the physical property measurements.

The sintered samples of $CsW_2O_6$ were synthesized by sintering the powder samples [26] for 10 min at 773 K in a spark plasma sintering furnace (SPS Syntex Inc., Japan) [30,31]. $CsW_2O_6$ decomposes when it is sintered via conventional solid-state reactions. By using spark plasma sintering, which can rapidly sinter a sample, the $CsW_2O_6$ powder can be sintered before it decomposes.

## 3. Electronic and structural properties of $CsW_2O_6$

3-1 Electronic properties

Figure 3 shows the temperature dependence of the electrical resistivity of $CsW_2O_6$ and $CsW_{2-\delta}O_6$ single crystals and a $CsW_2O_6$ sintered sample. The electrical resistivity ρ of $CsW_{2-\delta}O_6$ monotonically increases with decreasing temperature below room temperature. In contrast, ρ of $CsW_2O_6$ sharply increases at $T_t$ = 215 K for both single-crystalline and sintered samples [30]. As shown in Fig. 4, the magnetic susceptibility $\chi$ of $CsW_{2-\delta}O_6$ does not exhibit an anomaly, whereas $\chi$ for $CsW_2O_6$ shows a sharp decrease at $T_t$ with decreasing temperature, which is consistent with that of a polycrystalline sample [27]. The heat capacity divided by the temperature, $C/T$, of the $CsW_2O_6$ sintered sample exhibits a sharp peak at $T_t$, as shown in Fig. 5. A temperature hysteresis can be observed in the $\chi$ data at $T_t$, as shown in the inset of Fig. 4, indicating that a first-order phase transition occurs at $T_t$ in $CsW_2O_6$. As seen in the inset of Fig. 4, a sintered sample showed a sharper transition at $T_t$ than that of a single crystal, although the cause is unknown. The phases above and below $T_t$ are named Phases I and II, respectively.

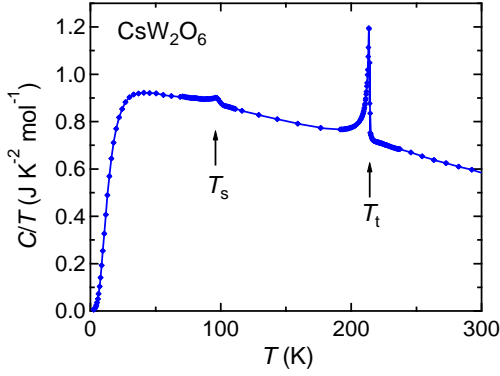

Figure 5. Temperature dependence of heat capacity divided by temperature of a sintered sample of $CsW_2O_6$. The data were obtained from Ref. 30. © 2020 The Authors.

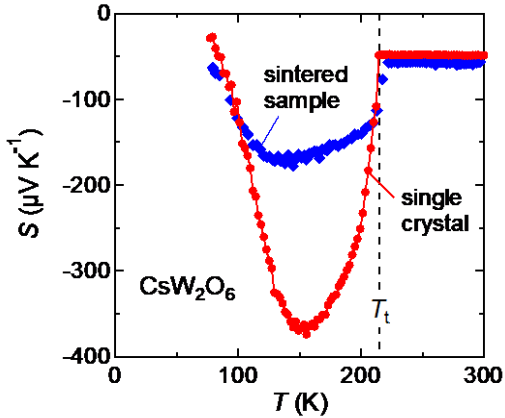

Figure 6. Temperature dependence of Seebeck coefficient of single-crystalline and sintered samples of $CsW_2O_6$. The data were obtained from Ref. 31. © 2020 The Physical Society of Japan.

A Fermi edge was observed in the valence-band photoemission spectra of the sintered samples and thin films in Phase I, indicating the presence of Fermi surfaces in this phase [29,32]. Therefore, Phase I is expected to exhibit metallic properties. However, Fig. 3 shows that $\rho$ weakly increases with decreasing temperature in Phase I for both single-crystalline and sintered samples, suggesting that the electronic state of Phase I below room temperature is not a simple metal. Particularly, $\rho$ of the single crystal is 3 mΩ cm at 300 K, which is a significantly higher value than that for a usual metal. As shown in Fig. 6, Seebeck coefficient $S$ of the $CsW_2O_6$ single crystal is almost constant at approximately $-50$ μV K$^{-1}$ in Phase I [31]. The negative value of $S$ is consistent with the electron configuration of $t_{2g}^{0.5}$, whereas the constant value of $S$, which varies from the case of $S \propto T$ for a typical metal, suggests that the electrical conduction in Phase I is represented by a hopping picture rather than an itinerant picture of coherent electrons. The $\chi$ in Phase I is not constant but considerably decreases with decreasing temperature, which is inconsistent with the Pauli paramagnetism.

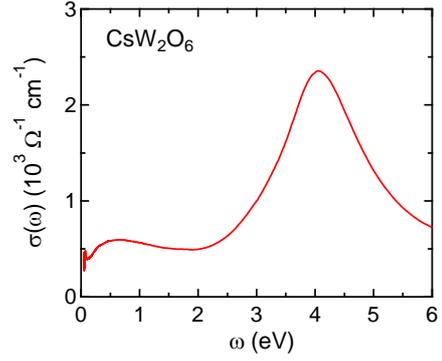

Figure 7. Optical conductivity of a single crystal of $CsW_2O_6$ at room temperature, as deduced from the reflectivity data using the Kramers-Kronig transformation. The Figure was adapted from Ref. 30. © 2020 The Authors.

The nonmetallic behavior in Phase I appeared more directly in the optical conductivity data [30]. As shown in Fig. 7, the optical conductivity $\sigma(\omega)$ of the $CsW_2O_6$ single crystal measured at room temperature shows a broad maximum at around $\hbar\omega = 0.6$ eV, and the extrapolated value to $\omega = 0$ is in good agreement with $\rho = 3$ mΩ cm at room temperature. This indicates the lack of or a negligible amount of Drude contribution, and the conducting electrons are trapped by something with an energy scale of 0.6 eV, resulting in the loss of coherency. The absence of a peak in the far-infrared region suggests that this localization is not due to disorder but might be reminiscent of the spectra of the lightly carrier-doped Mott insulators [33–35]. As discussed in Section 4-1, many pyrochlore oxides that exhibit a phase transition to an insulating state show a similar feature of the optical conductivity spectra, suggesting that this feature is not unique to $CsW_2O_6$ but is common for pyrochlore oxides. In addition, pair distribution function analysis of the synchrotron XRD data of Phase I indicated the presence of a local structure that differs from the $W_3$ trimer formation in Phase II, which is discussed subsequently [36]. This may be related to the nonmetallic features observed in Phase I.

In Phase II, $\rho$ exhibits semiconducting behavior as it increases exponentially with decreasing temperature. The Arrhenius plot of the electrical conductivity, shown in the inset of Fig. 3, indicates that both the single-crystalline and sintered samples exhibit thermally activated behavior in the low-temperature region. The linear fits yielded an activation energy $E_a$ of approximately 90 meV for both samples [31], which was consistent with $E_a$ = 92 meV for a polycrystalline sample [27]. These results suggest that Phase II is a semiconductor with a band gap corresponding to these $E_a$ values. The valence-band photoemission spectra measured using hard X-rays showed a shift of approximately 0.2 eV from 300 K (Phase I) to 180 K (Phase II), suggesting the opening of the energy gap of $\Delta$ = 0.2 eV at $E_F$ [32]. These $E_a$ and $\Delta$ values satisfy the relation of $2E_a \sim \Delta$, suggesting that the $E_a$ and $\Delta$ have been properly evaluated. The absolute value of $S$ also sharply increased below $T_t$, indicating that the number of conducting carriers in Phase II is significantly smaller than that in Phase I [31].

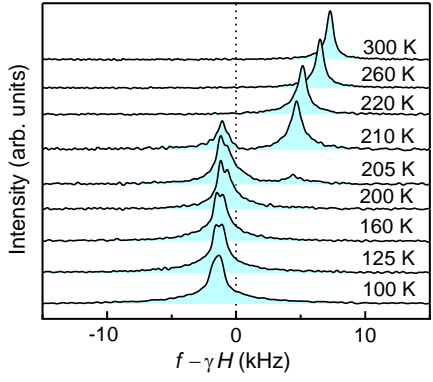

Figure 8. Temperature dependence of the $^{133}$Cs-NMR spectra measured in a magnetic field of 8 T applied along [001]. The Figure was adapted from Ref. 30. © 2020 The Authors.

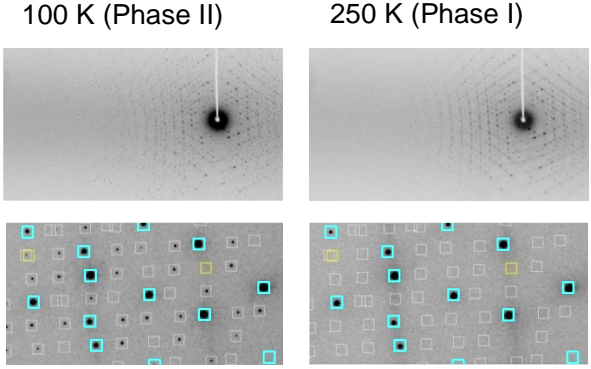

Figure 9. Single-crystal XRD patterns of a single crystal of CsW$_2$O$_6$ measured at 250 K (Phase I, right) and 100 K (Phase II, left). The pale blue and white squares indicate the positions of allowed and forbidden reflections, respectively, for the $Fd\bar{3}m$ space group. Figures were adapted from Ref. 30. © 2020 The Authors.

The magnetic susceptibility in Phase II is considerably smaller than that in Phase I, which is not due to an antiferromagnetic order but a result of spin singlet formation.

Table I. Crystallographic parameters for Phase I (250 K) of CsW$_2$O$_6$. The data were obtained from Ref. 30. © 2020 The Authors.

| Wyckoff position | | $x$ | $y$ | $z$ | $U_{eq}$ |
|---|---|---|---|---|---|
| W | 16$c$ | 0 | 0 | 0 | 0.00687(2) |
| Cs | 8$b$ | 3/8 | 3/8 | 3/8 | 0.02232(6) |
| O | 48$f$ | 0.06041(10) | 3/8 | 3/8 | 0.01064(10) |

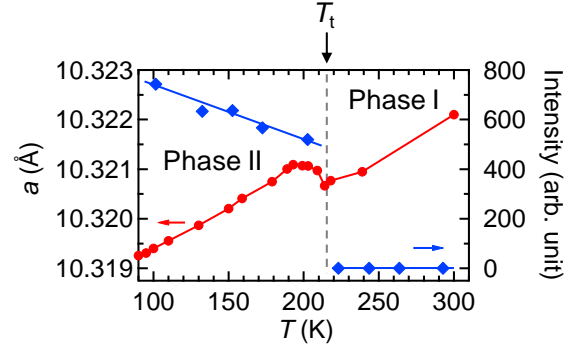

Figure 10. Temperature dependences of the lattice constant of CsW$_2$O$_6$ determined based on the single-crystal XRD data and the intensity of the (−2, −5, −3) reflection of the single crystal XRD data. The Figure was adapted from Ref. 30. © 2020 The Authors.

In the $^{133}$Cs-NMR spectra shown in Fig. 8, the peak shifts discontinuously between Phases I and II, but the line widths in Phase II do not show any significant broadening compared to those in Phase I, indicating the absence of an antiferromagnetic order in Phase II [30].

The $C/T$ data of CsW$_2$O$_6$ shown in Fig. 5 exhibit a small and distinct peak at $T_s$ = 90 K, in addition to a large peak at $T_t$ = 215 K. A structural phase transition occurs at $T_s$, which is discussed in Section 3-4, and the temperature region below $T_s$ was named Phase III. One of the characteristic features of this phase transition is that the electronic properties exhibit almost no anomalies at $T_s$, unlike in the case of the $T_t$ transition. For example, as shown in Fig. 4, no anomaly appeared in χ data at $T_s$ = 90 K. As discussed in Section 3-4, the phase transition at $T_s$ is most likely a pure structural transition resolving the structural instability of the β-pyrochlore structure.

3-2 Crystal structure of Phase II

Figure 9 shows the single-crystal XRD patterns of CsW$_2$O$_6$ measured at 250 K (Phase I) and 100 K (Phase II). All single-crystal XRD data presented in this section were measured at BL02B1 at SPring-8, Hyogo, Japan. The diffraction spots at 250 K were indexed on the basis of a cubic cell of $a$ = 10.321023(7) Å with the $Fd\bar{3}m$ space group, which were consistent with the results for the powder samples [26,30]. The crystallographic parameters at 250 K in Phase I are listed in Table I. The lattice parameter $a$ is larger than $a$ = 10.27190(10) Å for CsW$_{2-\delta}$O$_6$, reflecting the larger W composition in CsW$_2$O$_6$. In contrast, many diffraction spots appear in the diffraction patterns measured at 100 K, as shown in the panel on the left in Fig. 9. These diffraction spots were indexed on the basis of a cubic unit cell with $P2_13$ space group, which has approximately the same size as that of the cubic unit cell in Phase I. As in the case of the (−2, −5, −3) reflection shown in Fig. 10, which is forbidden and allowed in the $Fd\bar{3}m$ and $P2_13$ space groups, respectively, these reflections appeared upon entering Phase II. Their intensities increase discontinuously at $T_t$, reflecting a first-order phase transition. This result indicates that a rare structural transition that preserves cubic symmetry occurs at $T_t$. The crystallographic parameters at 100 K in Phase II are listed in Table II.

Table II. Crystallographic parameters for Phase II (100 K) of $CsW_2O_6$. The data were obtained from Ref. 30. © 2020 The Authors.

| Wyckoff position | $x$ | $y$ | $z$ | $U_{eq}$ |
|---|---|---|---|---|
| W(1) 4a | 0.378453(5) | 0.378453(5) | 0.378453(5) | 0.00286(1) |
| W(2) 12b | 0.625217(6) | 0.374558(6) | 0.628481(6) | 0.00356(1) |
| Cs(1) 4a | 0.243565(9) | 0.256435(9) | 0.743565(9) | 0.00925(2) |
| Cs(2) 4a | 0.500231(6) | 0.000231(6) | 0.499769(6) | 0.00914(3) |
| O(1) 12b | 0.56046(12) | 0.24962(7) | 0.75045(8) | 0.01268(18) |
| O(2) 12b | 0.50125(6) | −0.00598(6) | 0.81972(9) | 0.00539(10) |
| O(3) 12b | 0.50121(6) | 0.31034(9) | 0.49035(7) | 0.00576(9) |
| O(4) 12b | 0.74789(6) | 0.25218(6) | 0.56776(10) | 0.00565(10) |

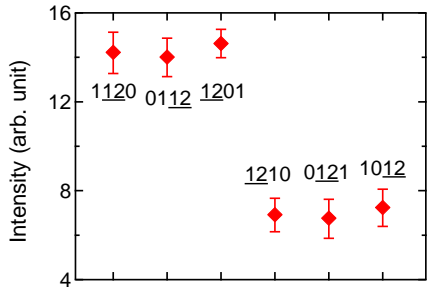

Figure 11. Intensities of the symmetrically equivalent reflections of $1\bar{2}10$ in the single-crystal XRD of 100 K (Phase II) of $CsW_2O_6$. The error bars indicate the standard deviation. The Figure was adapted from Ref. 30. © 2020 The Authors.

Because structural transitions that preserve the cubic symmetry are rare, confirming that Phase II is indeed cubic is important, despite not being an easy task. An important result supporting the cubic symmetry in Phase II is related to the intensities of the symmetrically equivalent reflections in the single-crystal XRD data. As shown in Fig. 11, the intensities of the $1\bar{1}20$, $01\bar{1}2$, and $\bar{1}201$ reflections and those of the $\bar{1}2\bar{1}0$, $0\bar{1}21$, and $10\bar{1}2$ reflections are identical within the uncertainties, respectively, indicating that Phase II belongs to the Laue class of $m\bar{3}$ [30]. In the case of $Fd\bar{3}m$ space group with the $m\bar{3}m$ Laue class, these six reflections have the same intensity. The fact that the intensity of the former three reflections differs from that of the latter three reflections indicates that the Laue class of Phase II is not $m\bar{3}m$. Moreover, as shown in Fig. 12(b), no anomaly is observed at $T_t$ in the temperature dependence of the full width at half maximum (FWHM) of the $\bar{1}197$ reflection, indicating that the cubic symmetry is maintained in Phase II. In contrast, the FWHM strongly increases below $T_s$, reflecting the monoclinic distortion in Phase III, which is discussed in Section 3-4.

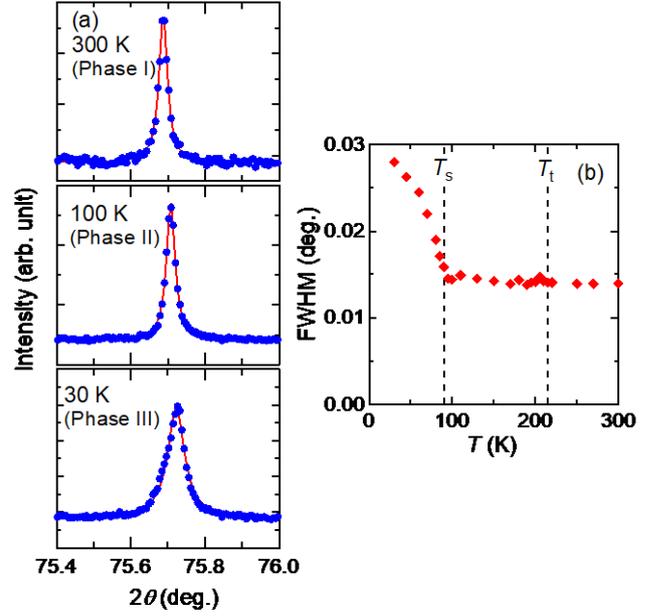

Figure 12. (a) Peak profiles of $\bar{1}197$ reflection (cubic unit cell) at 300 (Phase I), 100 (Phase II), and 30 K (Phase III) and (b) temperature dependence of the full width at half maximum of the $\bar{1}197$ reflection of the single crystal XRD data of $CsW_2O_6$. The solid curves in (a) show the fitting results to the Lorentzian function. The data were obtained from Ref. 30. © 2020 The Authors.

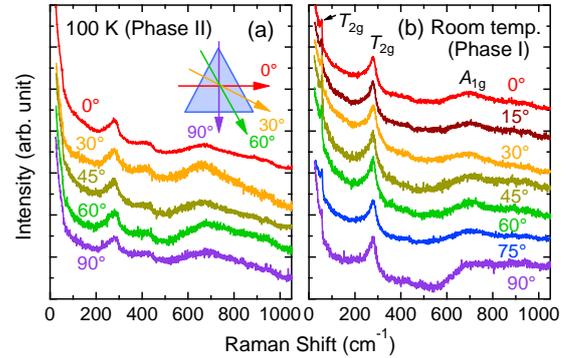

Figure 13. Polarization dependence of (111) Raman spectra of a single crystal of $CsW_2O_6$ measured at 100 K (a, Phase II) and room temperature (b, Phase I). Figures were adapted from Ref. 30. © 2020 The Authors.

Raman scattering is also sensitive to crystal symmetry, and aids in confirming the cubic symmetry in Phase II. Figure

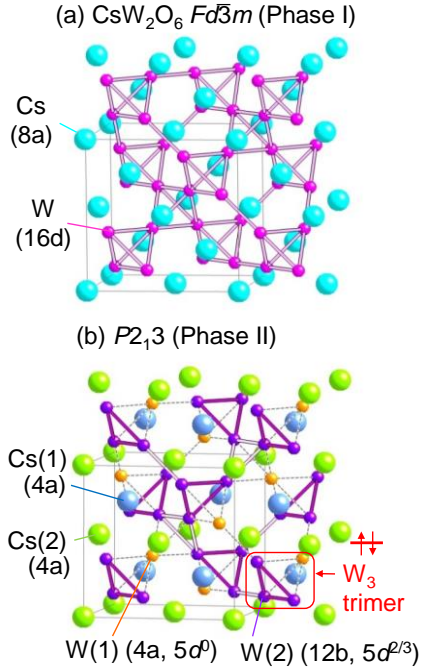

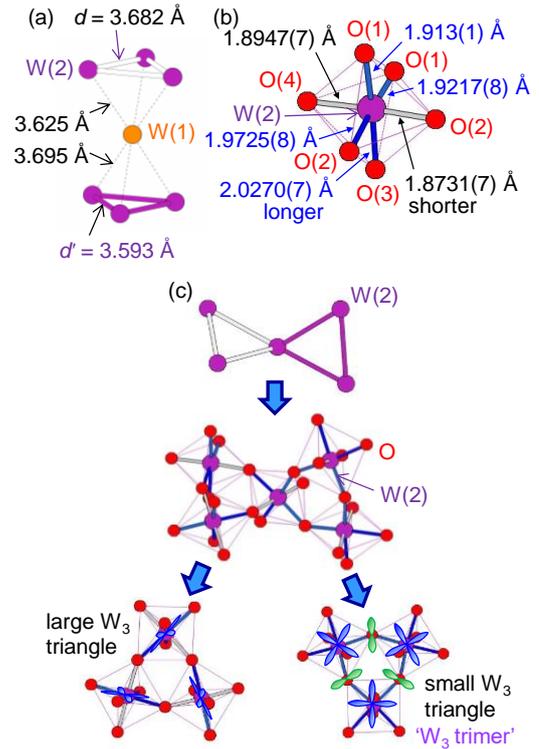

Figure 14. Crystal structures of the W and Cs sublattices at (a) Phase I ($T > 215$ K) and (b) Phase II ($90 < T < 215$ K). The purple and white triangles in (b) indicate the small and large $W_3$ triangles, respectively. Figures were adapted from Ref. 30. © 2020 The Authors.

Figure 15. (a) Distances between W atoms of $CsW_2O_6$ at 100 K (Phase II). (b) W–O bond lengths in a $W(2)O_6$ octahedron of $CsW_2O_6$ at 100 K (Phase II). (c) Schematic of the arrangement of occupied $5d$ orbitals in Phase II of $CsW_2O_6$. The lower-left and lower-right panels indicate the $W_3O_{15}$ units forming the large and small triangles made of W atoms, respectively. The small triangle corresponds to the $W_3$ trimer. In the lower right panel, $2p$ orbitals of the bridging oxygen atoms are also shown. In (a) and (c), the purple and white triangles indicate the small and large $W_3$ triangles, respectively. In (b) and (c), the blue and white W–O bonds indicate the four longer and two shorter bonds, respectively. A part of the Figures was obtained from Ref. 30. © 2020 The Authors.

13 shows the polarization dependence of the Raman spectra of the (111) surface of a single crystal of $CsW_2O_6$ measured at 100 K (Phase II) and room temperature (Phase I) [30]. As shown in Fig. 13(b), no angle dependence was observed in Phase I with the cubic symmetry. Similar to Phase I, there is also no angle dependence in the Raman spectrum of Phase II shown in Fig. 13(a), which was measured using the same crystal surface as the spectrum of Phase I shown in Fig. 13(b), supporting the presence of cubic symmetry in Phase II.

Here, the crystal structure of Phase II with the cubic $P2_13$ symmetry is discussed. In Phase II, Cs and W atoms occupied two sites each unlike in Phase I, where Cs and W atoms occupy one site each (Tables I and II). The Cs atoms, which formed a diamond structure in Phase I, formed a zinc-blende structure with the same numbers of Cs(1) and Cs(2) sites in Phase II. The presence of two Cs sites in Phase II is further verified by the $^{133}$Cs-NMR spectra shown in Fig. 8, which shows a single peak in Phase I and a slight split in Phase II, as clearly seen in the 125 K data.

The W atoms that form a pyrochlore structure in Phase I occupy the W(1) and W(2) sites in a ratio of 1:3 in Phase II. The W(2) atoms form a three-dimensional network of alternating small and large equilateral triangles that share their corners, as shown in Fig. 14(b). If these triangles are of the same size, the W(2) sublattice is equivalent to the hyperkagome structure [37], which indicates that the W(2) sites form a "breathing hyperkagome" structure [38]. As shown in Fig. 15(a), the difference in size between the large and small triangles formed by W(2) atoms, i.e., the breathing, is approximately 2% (the sizes of triangles are $d = 3.682$ Å and $d' = 3.593$ Å, respectively). As discussed in the next section, this breathing, although small at first glance, has a significant effect on the electronic properties of Phase II through the W $5d$ orbitals. Conversely, the W(1) atom is located in the middle of the small and large triangles formed by the W(2) atoms. As shown in Fig. 15(a), the distance between the W(1) atom and large triangle composed of W(2) atoms (white triangle) is approximately 2% shorter than that between the W(1) atom and the small triangle (purple triangle). Thus, the W(1) atom is farther from the small triangle and closer to the large triangle.

Because Cs and W atoms occupy two different sites each in Phase II, their valences cannot be uniquely determined. Here, their valences are discussed based on the bond valence sum (BVS) calculated using the interatomic distances determined from the single crystal XRD data [30,39]. First, the BVS for Cs is 1.00 at 250 K in Phase I and 1.03 and 1.00 for Cs(1) and Cs(2) at 100 K in Phase II, respectively. This result is consistent with the expectation that the valence of Cs is always 1+ in oxides, indicating that the crystallographic parameters used for the analysis were appropriately determined. In contrast, note that the BVS analysis of W is reliable for the $W^{6+}$ case because of the data accumulated from a large number of materials comprising $W^{6+}$, whereas it is less reliable for W atoms with a valence smaller than 6+ because the fewer materials with such valences are available. At 250 K, in Phase I, the BVS for W is 5.82; a value that is significantly less than 6 indicates that the W atoms are indeed reduced from $W^{6+}$, but the obtained value deviates by approximately 0.3 from the expected value of 5.5+. This gap is probably due to the aforementioned unreliability. At 100 K in Phase II, the BVS for the W(1) atoms was 6.07, indicating that the W(1) atom was $W^{6+}$ without $5d$ electrons. In contrast, the BVS of the W(2) site was 5.79 at 100 K. This value indicates that the W(2) atoms have a valence lower than 6+, unlike the W(1) atoms. However, the valence of the W(2) atoms cannot be quantitatively determined using BVS analysis alone. Considering the valences of the other atoms, i.e., $Cs^+$, $W(1)^{6+}$, and $O^{2-}$, and the charge balance, the W(2) atom is most likely $W^{5.33+}$ with two-third $5d$ electrons.

3-3 Electronic state in Phase II

As discussed in Section 3-1, $CsW_2O_6$ is a nonmagnetic insulator in Phase II. On the other hand, as discussed in Section 3-2, the crystal symmetry of Phase II is cubic $P2_13$, where W atoms occupy two sites and W(2) atoms, which are most likely to have two-third $5d$ electrons per atom, form a breathing hyperkagome structure. Let us consider the mechanism by which the nonmagnetic insulating state is achieved in Phase II. This is a unique self-organization of $5d$ electrons involving all spin, orbital, and charge degrees of freedom. Because the W(1) atom has no $5d$ electrons, the W(2) atoms play an important role. In the breathing hyperkagome structure of the W(2) atoms, the difference in size between the large and small $W_3$ triangles is small at approximately 2%. However, considering the W $5d$ orbital state, an essential difference exists between the small and large triangles.

The $5d$ electrons of the W(2) atoms are accommodated in the most stable of the five $5d$ orbitals. As shown in Fig. 15(b), the four longer W–O bonds in the six bonds in a $W(2)O_6$ octahedron (blue bonds), which are approximately 3–8% longer than the two shorter bonds (white bonds), comparable to the Jahn–Teller distortion in $t_{2g}$ electrons systems, exist in a plane. This suggests that the most stable $5d$ orbital occupied by electrons lies in this plane. In reality, this orbital is marginally hybridized with other orbitals, because the O–W–O angles are not 90°. As shown in Fig. 15(c), the relative arrangement of this orbital completely varies for the small and large $W_3$ triangles. In the small $W_3$ triangle, which is shown in the lower-right part of Fig. 15 (c), a large overlap between the occupied $5d$ orbitals via $2p$ orbitals of the bridging oxygen atoms is observed. In contrast, minimal overlap is observed in the large $W_3$ triangle, as shown in the lower-left part of Fig. 15(c). Here, the small and large $W_3$ triangles are alternately connected, as shown in Fig. 14(b); that is, a small triangle is always surrounded by large triangles. This situation indicates that the $5d$ electrons of a W(2) atom are confined in a small $W_3$ triangle composed of three W(2) atoms, which can be understood as the formation of a $W_3$ trimer or "molecule" on the breathing hyperkagome structure. According to the discussion in Section 3-2, two $5d$ electrons exist in this $W_3$ molecule (2/3 × 3 = 2), which doubly occupy a bonding orbital formed on this "molecule," resulting in the nonmagnetic insulating nature of Phase II.

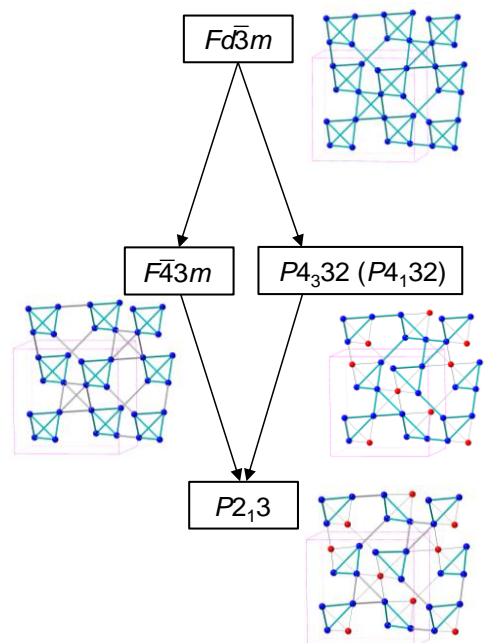

Figure 16. Group–subgroup relation between $Fd\bar{3}m$ and $P2_13$. A pyrochlore structure for each space group is also shown.

Thus, although the situation involving multiple degrees of freedoms in Phase II is complex, when considering the crystal symmetry, they can be understood as two events occurring simultaneously. Figure 16 shows the group–subgroup relationship from $Fd\bar{3}m$ in Phase I to $P2_13$ in Phase II. Because only the 8b (Cs), 16d (W), and 48f (O) sites are occupied by atoms in Phase I, only two possible routes exists from $Fd\bar{3}m$ to $P2_13$ according to the group–subgroup relationship. The first is via $F\bar{4}3m$, as shown on the left side of Fig. 16. In this case, 8b site splits into 4a and 4b sites from $Fd\bar{3}m$ to $F\bar{4}3m$, corresponding to the breathing of the pyrochlore structure [38,40]. From $F\bar{4}3m$ to $P2_13$, a change in centering from face-centered cubic to simple cubic struc-ture occurred, accompanied by the splitting of 16e site into 4a and 12b sites, corresponding to the hyperkagome order with a 1:3 ratio. The other route is via $P4_332$ ($P4_132$), as shown on the right side of Fig. 16. In this case, from $Fd\bar{3}m$ to $P4_332$, hyperkagome-type symmetry lowering occurs, where the centering changes from the face-centered cubic to simple cubic structure [37]. During the subsequent change from $P4_332$ to $P2_13$, breathing occurs in the hyperkagome structure.

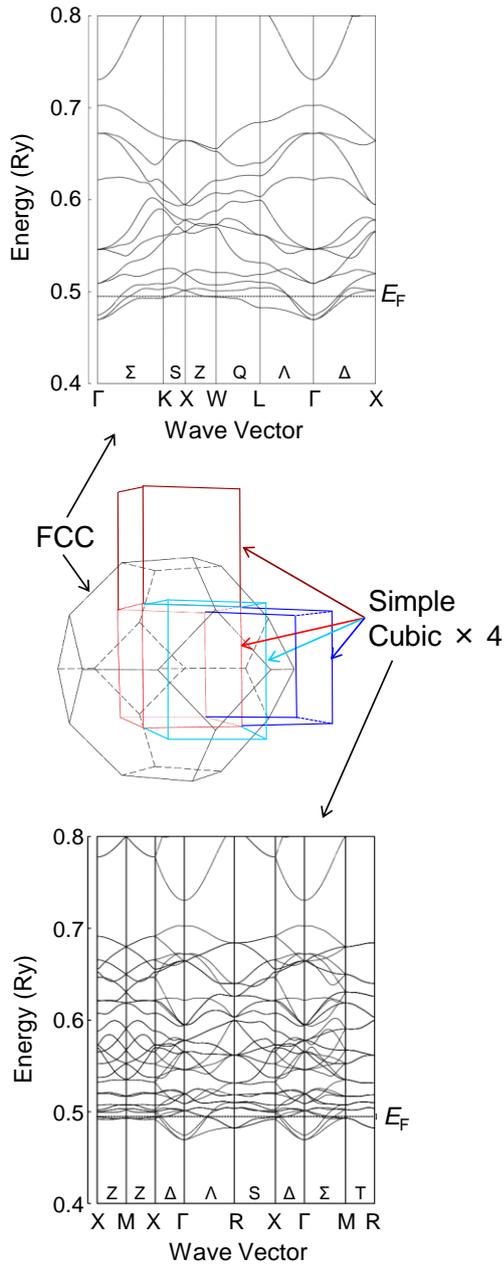

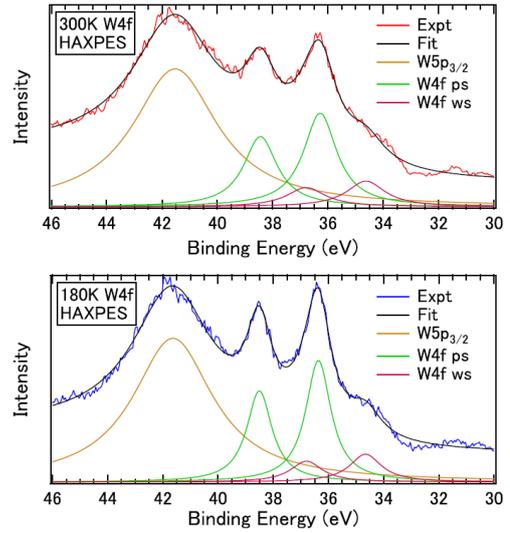

Figure 18. W $4f$ HAXPES spectra of $CsW_2O_6$ obtained at 300 K (upper, Phase I) and 180 K (lower, Phase II). The spectra are fitted to Lorentzian functions. The solid curves represent the fitted results and the Lorentzian components. In both panels, ps and ws denote poorly screened and well screened, respectively. Figures were adapted from Ref. 32. © 2022 American Physical Society.

Figure 17. Electronic structures of $CsW_2O_6$ in Phase I calculated with spin–orbit coupling. The top panel is calculated based on the Brillouin zone of the face-centered cubic lattice, which is shown by black lines in the middle panel. In the bottom panel, four band structures that are calculated based on the red, pale blue, blue, and brown simple cubic cells overlap in the graph. A part of the Figures were adapted from Ref. 30. © 2020 The Authors.

In summary, the group–subgroup relationship indicates that two types of symmetry changes, i.e., hyperkagome order and breathing, occur simultaneously during the phase transition from Phase I to Phase II of $CsW_2O_6$.

In these symmetry changes, the hyperkagome order is related to the electronic instability of Phase I, which can be interpreted as a "three-dimensional nesting" or "three-dimensional Peierls transition". The top panel of Fig. 17 shows the band structure of Phase I and the bottom panel shows four overlapping band structures, which are illustrated after the parallel shifts corresponding to the hyperkagome-type symmetry lowering, i.e., the change in centering from face-centered cubic to simple cubic structure [30]. As seen in the bottom panel of Fig. 17, the electronic bands intersect with each other near all points where the bands cross $E_F$, indicating that the Fermi surfaces are well-nested, corresponding to the parallel shift from face-centered cubic to simple cubic. In this case, the electronic system gains a large amount of energy when structural changes occur. This situation can be named "three-dimensional nesting." The Fermi surfaces of Phase I as reported in Ref. [28] are also well nested. They exhibit triple q Peierls instability, where three charge density waves (CDWs) with $\mathbf{q}_1 = (2\pi/a, 0, 0)$, $\mathbf{q}_2 = (0, 2\pi/a, 0)$, and $\mathbf{q}_3 = (0, 0, 2\pi/a)$ simultaneously condensate. Strong spin–orbit coupling plays an important role in achieving such a three-dimensional nesting. However, several limitations should be noted. The first is the incoherent metallic state in Phase I, as discussed in Section 3-1. The presence of a coherent metallic state may be important for determining whether such electronic instability in the band picture can be a driving force for phase transition. Second, such a hyperkagome order alone does not open an energy gap at $E_F$, according to first-principles calculations [28]. These results imply that the concept of three-dimensional nesting alone cannot explain the phase transition at $T_t$.

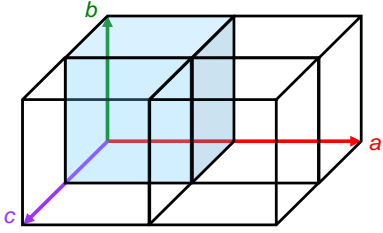

Figure 19. Unit cell of Phase III determined by the structural analysis of the single crystal XRD data. The unit cell of Phase III is 2 × 1 × 2 of that of Phase II, which is indicated as the blue shaded region. The Figure was adapted from Ref. 30. © 2020 The Authors.

Another important point to note regarding hyperkagome-type symmetry lowering is that the charge order of the W atoms is not obligatory. In the case of $P4_332$, the W atoms occupy 4b and 12d sites, and their positions are (5/8, 5/8, 5/8) and (1/8, $x$, $-x+1/4$), respectively. Therefore, symmetry lowering can be achieved by shifting the atomic positions without a difference in valence. The same is true for $P2_13$. Indeed, the HAXPES spectra shown in Fig. 18 suggest the absence of charge order of the W atoms in Phase II [32]. This point is be discussed in Section 4.2 in comparison with other materials.

Breathing, which is another symmetry-lowering process at $T_t$, is responsible for the formation of $W_3$ trimers through bond alternation in a hyperkagome structure. It would be interesting to observe what would happen if only breathing occurred at $T_t$. In this case, the $F\bar{4}3m$ symmetry would be realized, where a $W_4$ tetramer with a regular tetrahedron shape is formed instead of the $W_3$ trimer, as shown in the figure on the left in the middle row of Fig. 16. However, in this $F\bar{4}3m$ case, the energy gain is smaller than that in the case of $P2_13$, where both breathing and hyperkagome-type symmetry lowering occur, because the Jahn-Teller-like mechanism does not work in the $F\bar{4}3m$ case. In other words, $CsW_2O_6$ exhibits electronic instability that causes the hyperkagome order in Phase I and gains more energy through both hyperkagome order and breathing.

3-4 Phase III

$CsW_2O_6$ shows a phase transition from Phase II to Phase III at $T_s$ = 90 K [30]. As shown in Fig. 5, a small and distinct peak is observed in the $C/T$ data at 90 K, which corresponds to an entropy change of ~0.4 J K$^{-1}$ mol$^{-1}$. Further, as shown in Fig. 12, the FWHM of the 11$\overline{9}$7 reflection in the single-crystal XRD data suddenly increases below 90 K, which reflects the symmetry lowering from the cubic symmetry. In the cubic case, this reflection appears as a single peak. However, when the crystal symmetry is lower than cubic, the peak splits into multiple peaks, resulting in an increase in the FWHM. Although the structural parameters have not yet been determined owing to the complex crystal structure, the crystal symmetry of Phase III is determined to be a monoclinic symmetry with the $P2_1$ space group with a four-times larger unit cell than that in Phase II (Fig. 19). The increase in the FWHM of the 1197 peak reflects this monoclinic distortion.

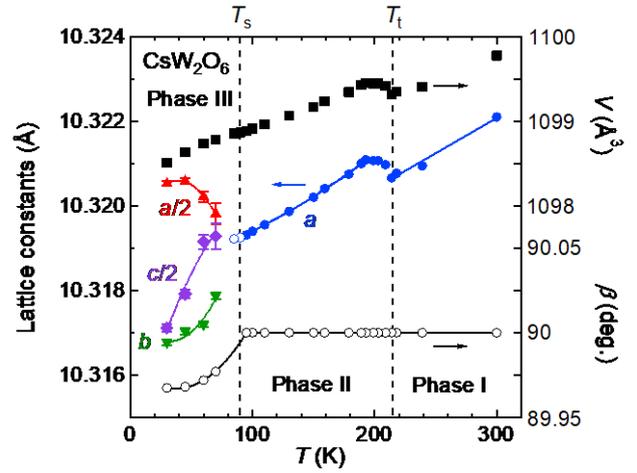

Figure 20. Temperature dependence of the lattice constants determined by the Rietveld analyses of the powder XRD data. The data were obtained from Ref. 30. © 2020 The Authors.

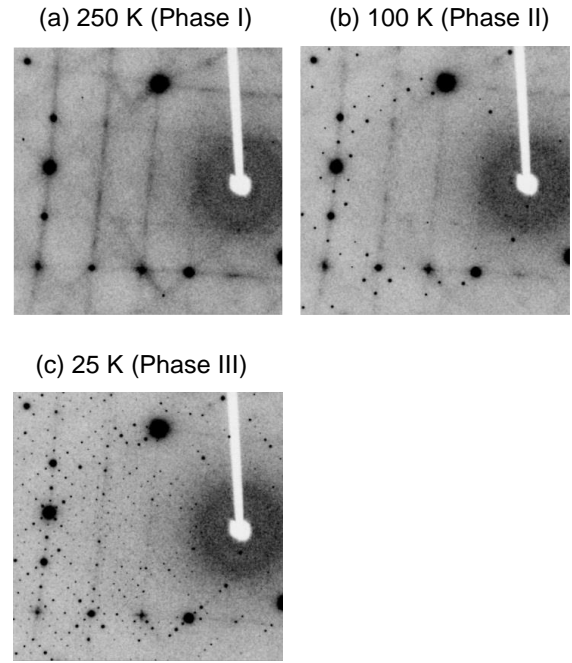

Figure 21. Single crystal XRD patterns taken at 250 (a, Phase I), 100 (b, Phase II), and 25 K (c, Phase III), where diffuse scatterings are emphasized. Since the intensity of diffuse scattering is much weaker than those of Bragg reflections, the observed diffuse scattering has no effect on the crystal-structure refinement. Figures were adapted from Ref. 30. © 2020 The Authors.

As can be observed from the temperature dependence of the lattice parameters shown in Fig. 20, the structural distortion in Phase III is small. Moreover, the anomalies associated with the phase transition at $T_s$ were hardly observed in the electronic properties, such as in $\chi$ shown in Fig. 4. These results suggest that the phase transition at $T_s$ is not electronic, unlike that at $T_t$. On the other hand, as shown in Fig. 21(c), the single-crystal XRD pattern of Phase III shows diffuse scattering at the position connecting the strong superlattice reflections appearing in Phase III. This diffuse scattering appears at the same positions in all the phases, as shown in Fig. 21, and follows an extinction rule that appears only at $h + k = 4n$, suggesting that the phase transition at $T_s$ and the observed diffuse scattering have a common origin. The same diffuse scattering also appears in $CsW_{1.835}O_6$ and $CsTi_{0.5}W_{1.5}O_6$, which are isostructural to $CsW_2O_6$ but do not have $5d$ electrons [30,41]. This suggests that the structural change from Phase II to Phase III and the diffuse scattering are induced by the structural instability of the β-pyrochlore oxides itself. Conversely, the fact that this structural instability is maintained down to 90 K indicates that the transition between Phases I and II is purely electronically driven.

## 4. Discussion

### 4-1 "Metallic" pyrochlore oxides

In this chapter, $CsW_2O_6$ is compared with various materials based on several perspectives, which highlights the unique properties and unsolved issues of $CsW_2O_6$, and provides a bird's-eye view of materials that share the characteristic features of $CsW_2O_6$. Table III lists the pyrochlore oxides that are metallic or exhibit high electrical conductivity even though $d\rho/dT < 0$ at room temperature. Many electrically conductive pyrochlore oxides exist, particularly $4d$ and $5d$ oxides. In terms of the $d$-electron configuration, high electrical conduction is achieved in a wide range from $d^{0.5}$ to $d^5$, where the $t_{2g}$ orbitals are partially occupied by $d$ electrons of transition metal atoms.

A characteristic feature of such pyrochlore oxides is that almost all exhibit a phase transition or an anomaly below room temperature. Few materials, limited to $Bi_2Ru_2O_7$, $Pb_2Ru_2O_{7-\delta}$, and $Pb_2Ir_2O_{7-\delta}$, remain in a simple paramagnetic metal down to the ground state [42–44]. $Cd_2Re_2O_7$ and $AOs_2O_6$ (A = K, Rb, and Cs) exhibit superconducting transitions at low temperatures [2,4,21–23]. Among these superconducting pyrochlore oxides, $Cd_2Re_2O_7$ and $KOs_2O_6$ show structural transitions [21, 45–50], as discussed subsequently.

Many of other pyrochlore oxides listed in Table III exhibit a magnetic order below room temperature. $Tl_2Mn_2O_7$, $Sm_2Mo_2O_7$, and $Nd_2Mo_2O_7$ exhibit ferromagnetic ordering. They exhibit interesting phenomena caused by the interplay between magnetism and conducting electrons, such as giant negative magnetoresistance and the anomalous Hall effect [51–54]. $Hg_2Os_2O_7$ and $Cd_2Ru_2O_7$ show an antiferromagnetic order [55–59]. $Ca_2Ru_2O_7$ and $Ca_2Ir_2O_7$ exhibit a spin-glass-like behavior with the temperature hysteresis of $\chi$ [60,61]. In $Ca_2Ru_2O_7$, this behavior was interpreted as a "frozen spin liquid" [62], which may be caused by oxygen deficiencies at the O′ site inherent in the Ca-based pyrochlore oxides.

In contrast to these pyrochlore oxides with metallic or electrically conducting ground states, several pyrochlore oxides, such as $Cd_2Os_2O_7$, $Hg_2Ru_2O_7$, and $Nd_2Ir_2O_7$ exhibit an antiferromagnetic order accompanied by a metal-insulator transition, resulting in an insulating ground state [9,10,12,14–17,63,64]. The antiferromagnetic ordering temperatures are $T_N$ = 227, 107, and 33 K for $Cd_2Os_2O_7$, $Hg_2Ru_2O_7$, and $Nd_2Ir_2O_7$, respectively. The magnetic structures of $Cd_2Os_2O_7$ and $Nd_2Ir_2O_7$ are all-in-all-out, which does not cause a reduction in the crystal symmetry [9,10,12,64]. In $Hg_2Ru_2O_7$, the structural distortion associated with magnetic order is significantly small [15,65]. These results suggest that the metallic states in these pyrochlore oxides are easily lost because of changes in the surrounding environment. In fact, a metal-insulator transition in $Cd_2Os_2O_7$ was explained via the

Table III. Electrically conductive pyrochlore oxides. All materials with more electrons than $d^3$ have the low-spin states.

| filling | material | room temp. | ground state | ref |
|---|---|---|---|---|
| $5d^{0.5}$ | $CsW_2O_6$ | NM | I | [27,30] |
| $4d^2$ | $Ln_2Mo_2O_7$ (Ln = Gd, Sm, Nd) | M | M | [66,67] |
| $5d^2$ | $Cd_2Re_2O_7$ | M | SC | [21,22,68] |
| $5d^{2.5}$ | $AOs_2O_6$ (A = K, Rb, Cs) | M | SC | [2,4,23] |
| $5d^{2.75}$ | $Pb_2Re_2O_{6.75}$ | M | M | [69] |
| $3d^3$ | $Tl_2Mn_2O_7$ | M | M | [70] |
| $4d^3$ | $Cd_2Ru_2O_7$ | NM | NM | [58] |
| $4d^3$ | $Hg_2Ru_2O_7$ | M | I | [14,15] |
| $4d^3$ | $Ca_2Ru_2O_7$ | NM | NM | [60] |
| $5d^3$ | $Cd_2Os_2O_7$ | M | I | [13] |
| $5d^3$ | $Hg_2Os_2O_7$ | M | M | [55] |
| $4d^{3.5}$ | $Pb_2Ru_2O_{6.5}$ | M | M | [42] |
| $4d^4$ | $Bi_2Ru_2O_7$ | M | M | [42] |
| $4d^4$ | $Tl_2Ru_2O_7$ | NM | I | [18–20] |
| $5d^4$ | $A_2Ir_2O_7$ (A = Ca, Cd) | M | M | [61,71] |
| $5d^{4.45}$ | $Pb_2Ir_2O_{6.55}$ | M | M | [43,44] |
| $5d^5$ | $Pr_2Ir_2O_7$ | M | M | [72] |
| $5d^5$ | $Nd_2Ir_2O_7$ | M | I | [16,17] |
| $5d^5$ | $Ln_2Ir_2O_7$ (Ln = Sm, Eu, Gd, Tb, Ho, Dy) | NM | I | [16,17] |
| $5d^5$ | $Tl_2Ir_2O_7$ | M | M | [73] |

I: insulator, NM: nonmetal, M: metal, SC: superconductor

Lifshitz transition, in which a semimetallic overlap between the valence and conduction bands is lost owing to the magnetic order. Moreover, the magnetic order is accompanied by a transition to the insulating state depending on the subtle differences in the constituent elements, as shown by a comparison between $Hg_2Os_2O_7$ (metallic) and $Cd_2Os_2O_7$ (insulating).

$CsW_2O_6$ differs from these pyrochlore oxides in that the phase transition to the insulating state is not accompanied by magnetic order. Among the other pyrochlore oxides, only $Tl_2Ru_2O_7$ showed a similar transition. The stoichiometric $Tl_2Ru_2O_7$ synthesized under high oxygen pressure exhibits a metal-insulator transition at approximately 120 K [19,20]. The insulating phase has orthorhombic symmetry with the space group of $Pnma$ [20]. The $RuO_6$ octahedra are highly distorted in the insulating phase by approximately 6% and 11% at the two Ru sites, as in the case of Jahn–Teller distortion. As a result of the orbital order occurring with the Jahn–Teller distortion, Ru $4d$ electrons are confined in one-dimensional chains made of Ru atoms and form spin-singlet pairs on the chains, resulting in a nonmagnetic insulating state [20,74]. This is similar to the nonmagnetic $W_3$ trimer formation in $CsW_2O_6$. On the other hand, unlike $CsW_2O_6$, the Tl $6s$ electrons may be involved in the metal-insulator transition in $Tl_2Ru_2O_7$. Furthermore, the oxygen vacancy often occurs at the O′ site, which has a large effect on the electronic properties. Thus, $Tl_2Ru_2O_7$ shows an electronic phase transition similar to that of $CsW_2O_6$, but is more complex.

$CsW_2O_6$ also differs from other pyrochlore oxides in that it exhibits a phase transition to the insulating state, in which the W atoms have a half-integer valence. Pyrochlore systems with half-integer valences, including spinels, exhibit interesting features in terms of the charge order, which are discussed in Section 4.2.

$CsW_2O_6$ is compared with other electrically conducting pyrochlore oxides in terms of the structural distortion at the phase transition. For example, $Cd_2Re_2O_7$ exhibits multi-step structural phase transitions, with the first transition at 202 K, which are of interest in terms of the electric toroidal quadrupole [75–78]. $KOs_2O_6$ shows an isostructural phase transition preserving $Fd\bar{3}m$ symmetry at 7.5 K, which is in the superconducting phase [47–50]. The metal-insulator transition in $Hg_2Ru_2O_7$ is accompanied by a small structural distortion [15,65]. The structural changes at these transitions are small. In contrast, the phase transitions in $CsW_2O_6$ at $T_t$ and in $Tl_2Ru_2O_7$ are accompanied by large distortions of the $WO_6/RuO_6$ octahedra [20,30,74]. As in the case of spinels, which is discussed in the next section, this suggests that the orbital and/or charge order plays a central role in the phase transition, unlike in other pyrochlore oxides. In contrast, the structural distortion at the $T_s$ transition of $CsW_2O_6$ is significantly small, similar to that in several pyrochlore oxides discussed previously.

Next, the nonmetallic behavior in Phase I of $CsW_2O_6$ is examined. As discussed in Section 3-1, $CsW_2O_6$ single crystals show resistivity with $d\rho/dT < 0$ and the optical conductivity spectrum with a broad maximum at around 0.6 eV in Phase I. Similar behaviors were observed in several pyrochlore oxides exhibiting a metal-insulator transition. First, the "metallic" phase of $Tl_2Ru_2O_7$ exhibited almost temperature-independent ρ and an optical conductivity spectrum with a broad maximum at around 0.5 eV [19,20,74,79]. This maximum shows a strong temperature dependence that cannot be explained by interband transitions or polaron excitations. Rather, this behavior is similar to that of the mid-gap states in doped Mott insulators, as observed in doped $YVO_3$ [79–81]. The optical conductivity spectra of $Nd_2Ir_2O_7$ near the metal-insulator transition temperature showed a broad maximum at approximately 0.05–0.1 eV instead of a Drude-like increase [82]. Thus, the optical conductivity spectra of $CsW_2O_6$, $Tl_2Ru_2O_7$, and $Nd_2Ir_2O_7$ commonly exhibit the characteristic features of the metal-insulator transitions of strongly correlated Mott insulators. Because pyrochlore oxides have narrow $d$ bands due to the bent M–O–M bonds, they can exhibit moderately strong electron correlations for the $4d$ and $5d$ electron systems, which may be related to the aforementioned nonmetallic behaviors. In contrast, no peaks corresponding to the mid-gap state are observed in the optical conductivity spectra of $Cd_2Os_2O_7$ [83]. The metal-insulator transition of $Cd_2Os_2O_7$ was basically understood within the Lifshitz type unlike other pyrochlore oxides [12,83], although the spectral weight was affected by the electron correlation.

Thus, many electrically conducting pyrochlore oxides exist. Many of them show some kind of phase transition including the "metal"-insulator transition, but the structural changes associated with these phase transitions are often small or absent. Therefore, the phase transitions to the insulating state in $CsW_2O_6$ and $Tl_2Ru_2O_7$ are exceptional because they are accompanied by large structural distortions. They also differ from those in other pyrochlore oxides in that they are not associated with the magnetic order. In contrast, nonmetallic behavior above the "metal"-insulator transition is commonly observed in many pyrochlore oxides, including $CsW_2O_6$.

One reason for the exceptional properties of $CsW_2O_6$ is its small number of $5d$ electrons. As shown in Table III, the electrically conducting pyrochlore oxides are almost evenly distributed from $d^2$ to $d^5$; however, $CsW_2O_6$ is the only oxide with a smaller $d$-electron configuration. Some $d^1$ pyrochlore oxides, such as $Lu_2V_2O_7$, exist. However, they are all insulators in all measured temperature ranges [3]. For the $d^2$–$d^5$ electron configurations, various electronic phenomena are achieved, which are often related to magnetism. In the future, we expect to find a new material with fewer electrons than $d^2$, following $CsW_2O_6$, which will lead to the emergence of novel electronic phenomena that differ from those in existing pyrochlore oxides.

4-2 Charge order on a pyrochlore structure

Studies on electronic phase transitions in transition metal compounds in which transition metal atoms with half-integer valence form a pyrochlore structure have a long history spanning nearly a century. Magnetite $Fe_3O_4$ was reported to exhibit a metal-insulator transition with a charge order of Fe atoms at 119 K. This phase transition was later referred to as the Verwey transition [84,85]. Despite extensive research, a complete understanding of the Verwey transition including the mechanism has not yet been achieved. Magnetite $Fe_3O_4$ is an inverse spinel, where $Fe^{3+}$ atoms occupy the tetrahedral sites and Fe atoms with an average valence of 2.5+ occupy the octahedral sites, i.e., they are written as $Fe^{3+}Fe^{2.5+}_2O_4$. Below the Verwey transition, the $Fe^{2.5+}$ atoms that form a pyrochlore structure change into the charge-ordered states of $Fe^{2+}$ and $Fe^{3+}$. The question of how identical numbers of $Fe^{2+}$ and $Fe^{3+}$ are arranged in the pyrochlore structure has been researched [86,87]. To minimize the Coulomb energy, the total charge of each tetrahedron of the pyrochlore structure must be equal, which is called the Anderson condition. When placing $Fe^{2+}$ and $Fe^{3+}$ on a tetrahedron to satisfy the Anderson condition, the Coulomb energy can be minimized if $Fe^{2+}$ and $Fe^{3+}$ atoms are adjacent to each other to the maximum possible extent; however, they cannot be adjacent on every edge. $Fe^{2+}$ and $Fe^{3+}$ can be adjacent on four edges in a tetrahedron, but on the other two edges, Fe atoms with the same valence are adjacent. Each tetrahedron exhibits degeneracy in the placement of these two edges. Anderson pointed out that this macroscopic degeneracy significantly suppresses the temperature of the Verwey transition [88].

However, until the report on the charge order in $CsW_2O_6$, a charge order that satisfies the Anderson condition was not discovered. In addition to $Fe_3O_4$, certain materials, such as $CuIr_2S_4$ and $AlV_2O_4$, were observed to exhibit a charge order. However, a charge order violating the Anderson condition has always been achieved [89–93]. Notably, these are all spinel compounds in the $t_{2g}$ electron system. In spinels, the octahedra of anionic atoms, such as oxygen and sulfur, surrounding the transition metal atoms are connected by sharing their edges, unlike α- and β-pyrochlore oxides. The $t_{2g}$ orbitals of the transition metal atoms extend toward neighboring transition metal atoms. Therefore, the energy gain when a σ bond is formed between $t_{2g}$ orbitals is large [94,95]. Consequently, as the loss of Coulomb energy due to the violation of the Anderson condition can be fully compensated by the σ bond formation, a charge order with many "molecules" formed by the σ bonds is always chosen. The shape and density of the molecules depend on the materials used. In $Fe_3O_4$, various charge-ordering patterns, including the "trimeron," have been proposed [96]. In $CuIr_2S_4$, a charge order of $Ir^{3+}$ and $Ir^{4+}$ occurs as a metal-insulator transition at 230 K [90,91,97]. The insulating phase was reported to have a complex octamer made of $Ir^{4+}$ atoms [90], which was later understood as the $Ir_2$ dimer formation based on the "orbitally induced Peierls" mechanism [98]. In $AlV_2O_4$, the formation of a $V_7$ heptamer in the charge-ordered phase was suggested; however, more recently, the charge order has been understood by the formation of $V_3^{9+}$ trimers and $V_4^{8+}$ tetramers [92,99]. The Anderson condition is completely violated in this $V_4^{8+}$ tetramer, because this tetrahedron comprises only $V^{2+}$ atoms.

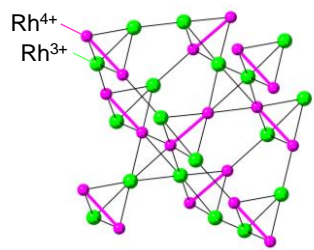

Figure 22. Rh sublattice in $LiRh_2O_4$ at 30 K [104]. The small and large spheres represent the $Rh^{4+}$ and $Rh^{3+}$ atoms, respectively. The thick solid lines between $Rh^{4+}$ atoms indicate the σ bond in a $Rh^{4+}_2$ dimer.

In contrast, the charge order in $CsW_2O_6$, discussed in Sections 3-2 and 3-3, satisfies the Anderson condition. As shown in Fig. 14(b), every $W_4$ tetrahedra forming a pyrochlore structure is equivalent because it consists of one W(1) and three W(2) atoms, indicating that the total charge of each tetrahedron is equal. As $CsW_2O_6$ has a β-pyrochlore structure, it gains a smaller energy owing to the bond formation than those in the spinels, because the $WO_6$ octahedra in $CsW_2O_6$ share their corners, which is disadvantageous for the formation of σ bonds. Consequently, the Coulomb energy becomes more important, and the charge order satisfying the Anderson condition becomes the most stable. However, this charge order is not a simple 1:1 ratio of W atoms with an integer valence. The nontrivial 1:3 ratio of $W^{6+}$ and $W^{5.33+}$ with a hyperkagome order, avoiding the geometrical frustration of the pyrochlore structure, is achieved in Phase II of $CsW_2O_6$. This hyperkagome-type order is a natural pattern that often appears in pyrochlore systems with a 1:3 ratio of two types of atoms, such as B-site ordered spinel oxides $Li_2BB'_3O_8$, B-site ordered Laves $A_2BB'_3$, and the up-up-up-down spin structure in the half magnetization plateau of Cr spinels [37,100–103]. However, $CsW_2O_6$ is unique in that this type of order is achieved by combining the fractional valence and molecular formation, which is a novel self-organization phenomenon of $d$ electrons based on a traditional problem in solid state physics.

More recently, the charge ordering pattern of spinel $LiRh_2O_4$ has been clarified, as shown in Fig. 22, where the charge order of $Rh^{3+}$ and $Rh^{4+}$ with the $Rh^{4+}_2$ dimer occurs while satisfying the Anderson condition [104]. $LiRh_2O_4$ has the same $d$-electron configuration as $CuIr_2S_4$ and is similar to $CuIr_2S_4$ in that the orbital order is involved in the charge order [105]. This result demonstrates the importance of further exploration of new materials for a full understanding of the charge order on pyrochlore structures and the discovery of novel related phenomena.

However, whether and how such a charge order occurs in $CsW_2O_6$ are important questions. In particular, as shown in Fig. 18, the results of the bulk-sensitive HAXPES do not support the charge order of $W^{6+}$ and $W^{5.33+}$ [32]. The W $4f$ HAXPES spectrum shows a main peak and a smaller shoulder on the lower binding energy side in both Phases I and II. because a lower binding energy corresponds to a lower valence, a 1:3 ratio of $W^{6+}$–$W^{5.33+}$ charge order should show a larger peak with a lower binding energy. However, this is not the case in the obtained spectrum. Instead, this shoulder peak can be explained by the screening effect, which often occurs in moderately correlated systems. This result contrasts with the photoemission spectra of $Fe_3O_4$, $CuIr_2S_4$, and $LiRh_2O_4$, where different charge states were observed for Fe, Ir, and Rh [97,106,107].

Considering the HAXPES results alone, the electronic phase transition at $T_t$ appears to be achieved by a CDW-like mechanism rather than by charge order. Conversely, as shown in Fig. 15(b), a large Jahn–Teller-like distortion exists in the $WO_6$ octahedra in Phase II [30], suggesting that a "molecule" picture is more appropriate for the phase transition at $T_t$. In this case, each atom gains energy, which is different from the CDW-like mechanism, in which electrons near $E_F$ gain energy owing to the small lattice distortion. Moreover, the results of the BVS analysis based on the synchrotron XRD data provide a strong evidence for a 1 : 3 ratio of $W^{6+}$ and W atoms with a significantly lower valence than 6+, i.e., the presence of charge order or charge disproportionation of W atoms in Phase II, although this is indirect evidence obtained though the position of the ligands [30]. We hope that future experimental and theoretical studies will elucidate how such a self-organization of $d$ electrons occurs and how is observed.

4-3 Equilateral-triangular trimer formation

The formation of equilateral-triangular trimers composed of W atoms, that is, $W_3$ molecules, is one of the characteristic features of Phase II of $CsW_2O_6$. Such equilateral-triangular trimers are often formed in V and Ti compounds. The most classic example is layered rock-salt $LiVO_2$, which exhibits a phase transition from a high-temperature nonmetallic phase to a low-temperature nonmagnetic insulating phase, accompanied by the formation of equilateral-triangular $V_3$ trimers at approximately 500 K [108–113]. In addition to this compound, triangular $V_3$ ($Ti_3$) trimers have been reported in various materials, such as $LiVS_2$ [114,115], $Na_2Ti_3Cl_8$ [116–118], and $Ba_2V_{13}O_{22}$ (marginally distorted) [119–121]. They are all composed of edge-shared $VX_6$ or $TiX_6$ octahedra, where the valences of V and Ti are 3+ and 2+, respectively. In these cases, they comprise two $3d$ electrons that occupy two of the $3d_{xy}$, $3d_{yz}$, and $3d_{zx}$ orbitals [113]. As shown in the panel on the right of Fig. 23, these orbitals extend toward the adjacent V/Ti atoms, forming σ bonds between $3d$ orbitals, as in the cases of the spinels discussed in Section 4-2. Two such σ bonds are formed on a V/Ti atom, forming a three-membered ring of V/Ti atoms with the three σ bonds. The six $3d$ electrons of the three V/Ti atoms completely occupy the bonding orbitals, resulting in a nonmagnetic insulating ground state. This is a typical cluster formation by the chemical bonds in solids.

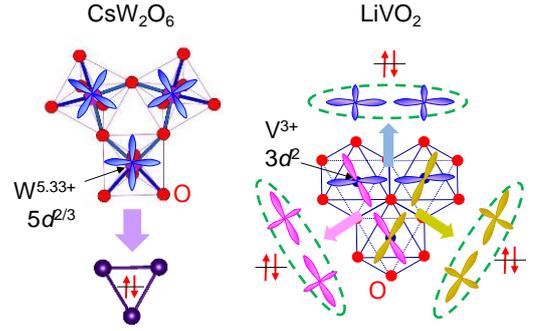

Figure 22. Trimer formation in $CsW_2O_6$ (left) and $LiVO_2$ (right).

The trimer formation in $CsW_2O_6$ is similar to that in the aforementioned V and Ti systems, as in the trimer is nonmagnetic and has an equilateral-triangular shape. However, it is different in that the trimer is formed by two $5d$ electrons. As shown in the panel on the light of Fig. 23, two $5d$ electrons (2/3 × 3 = 2) doubly occupy a molecular orbital across the three W atoms, which can be understood as a three-center-two-electron (3c2e) bond formation. The 3c2e bond is one of the most common forms of multi-centered bonds and often appears in electron-poor compounds such as borane [122]. Compared to a normal single bond, which is formed by two electrons donated by two atoms, the bond order in a 3c2e bond is lower. However, the number of atoms involved in the bond is larger. Because $CsW_2O_6$ is an electron-deficient compound, where the W atom has a half $5d$ electron on average, 3c2e bond formation is a natural choice. However, the 3c2e bond formation rarely occurs as a temperature-induced phase transition. This implies that the molecules consisting of the 3c2e bond are formed (dissociated) by lowering (increasing) the temperature on the solid. Recently, an electron-rich compound, RuP, was reported to exhibit a phase transition accompanied by a three-center-four-electron (3c4e) bond formation [123], indicating that the multi-centered bond formation is a universal phenomenon in solids.

Another interesting aspect of the 3c2e bond formation in $CsW_2O_6$ is its equilateral triangular shape. The 3c2e bonds usually have a bent shape, as in borane. Only the $H_3^+$ ion has an equilateral-triangular shape among the materials with 3c2e bonds. $H_3^+$ is an interstellar material that is unstable on Earth [124]. The equilateral-triangular shape of $H_3^+$ is related to the formation of a 3c2e bond via the $1s$ orbital of hydrogen. In $CsW_2O_6$, a pyrochlore structure most likely acted as a template, resulting in the equilateral-triangular $W_3$ molecule. This indicates that crystalline solids can produce novel types of molecules.

4-4 Cubic-cubic structural transition

Finally, we discuss the cubic-cubic structural transition in $CsW_2O_6$, which is also a characteristic feature of this compound. Perhaps the most famous example of a phase transition in which the symmetry of crystal-structure changes while maintaining its cubic symmetry is the α–γ transition in iron. Pure iron shows a phase transition from the ferrite phase with a body-centered cubic structure to the austenite phase with a face-centered cubic structure at 911 °C with increasing temperature [125]. Several examples of structural phase transitions that preserve the cubic symmetry are also known for order-disorder transitions in solids [126]. One is the order-disorder transition in ordered alloys such as $Cu_3Au$. $Cu_3Au$ has a face-centered cubic structure with the $Fm\bar{3}m$ space group at high temperatures, where the Cu and Au atoms randomly occupy each lattice point. At lower temperatures, the Cu and Au atoms regularly occupy the face-centered cubic structure, resulting in a $Cu_3Au$ structure with $Pm\bar{3}m$ symmetry. $C_{60}$ fullerene exhibits another type of order-disorder transition. At high temperatures, $C_{60}$ fullerene crystallizes in a face-centered cubic structure with $Fm\bar{3}m$ symmetry, in which the $C_{60}$ molecules rotate freely. With decreasing temperature, the rotation stops at 260 K, at which the crystal symmetry changes to cubic $Pa\bar{3}$ owing to the long-range order of the orientation of the $C_{60}$ molecules [127]. Thus, structural transitions accompanied by a large rearrangement of atomic positions often preserve the cubic symmetry.

In contrast, electronic phase transitions rarely appear as a cubic-cubic transition, such as the phase transition of $CsW_2O_6$ at $T_t$. In the cubic-cubic electronic phase transition, not only the complex self-organization of electrons as in the case of $CsW_2O_6$, but also various physical properties unique to cubic systems, such as multipole order and nontrivial electromagnetic responses, are expected to occur. $GaNb_4Se_8$ and $PrRu_4P_{12}$ are examples of electronic phase transitions that preserve the cubic symmetry [128,129]. $GaNb_4Se_8$ is a lacunar spinel, which is a spinel with half of the tetrahedral sites regularly missing; the crystal structure has the cubic $F\bar{4}3m$ space group, which is a subgroup of $Fd\bar{3}m$, as shown in Fig. 16. $GaNb_4Se_8$ is expected to be a "molecular" $J_{eff} = 3/2$ electron system, in which $J_{eff} = 3/2$ pseudospins are carried by the molecular orbital in a $Nb_4Se_4$ cluster [130]. This is in contrast to the skyrmion lattice magnets $GaV_4S_8$ and $GaV_4Se_8$, where spin-1/2 is carried by the $V_4S_4/V_4Se_4$ clusters [131,132]. $GaNb_4Se_8$ exhibits two phase transitions at $T_Q$ = 50 K and $T_m$ = 33 K. Below $T_m$, $GaNb_4Se_8$ has orthorhombic $P2_12_12_1$ symmetry and a nonmagnetic ground state. The phase transition at $T_Q$ is interesting in terms of the cubic-cubic transition. At $T_Q$, the crystal symmetry changes from cubic $F\bar{4}3m$ ($T > T_Q$) to cubic $P2_13$ ($T < T_Q$) [128,133]. This symmetry change is caused by a hyperkagome-type 1:3 charge order, similar to that in Phase II of $CsW_2O_6$ [133]. However, because the average valence of Nb in $GaNb_4Se_8$ is 3.25+, the 1:3 charge order in $GaNb_4Se_8$ is an integer valence of $Nb^{4+}$ and $Nb^{3+}$, which is in contrast to the case of $CsW_2O_6$ with a fractional valence ($W^{6+}$ and $W^{5.33+}$). $Nb^{3+}$ atoms ($4d^2$) form nonmagnetic $Nb_3$ trimers with an equilateral-triangular shape, similar to those in $LiVO_2$ and the other materials discussed in Section 4-3. In contrast, $Nb^{4+}$ atoms ($4d^1$) have a localized spin and are associated with a phase transition at $T_m$.

A filled skutterudite phosphide $PrRu_4P_{12}$ exhibits a metal-insulator transition at 63 K [129]. This phase transition is accompanied by a structural change from cubic $Im\bar{3}$ in the high-temperature metallic phase to cubic $Pm\bar{3}$ in the low-temperature insulating phase [134]. $PrRu_4P_{12}$ has well-nested Fermi surfaces in the metallic phase. This nesting corresponds to the change in centering from body-centered cubic to simple cubic [135], which is similar to the three-dimensional nesting in $CsW_2O_6$. The temperature evolution of the intensity of the superlattice reflections in $PrRu_4P_{12}$ indicates that the low-temperature phase was in a CDW state corresponding to this nesting [136]. Interestingly, however, $LaRu_4P_{12}$ with similar nesting but without $4f$ electrons, does not exhibit a phase transition [135,137]. In the low-temperature phase of $PrRu_4P_{12}$, antiferro-quadrupolar order is achieved, where the quadrupoles of the Pr $4f$ electrons are ordered in the CsCl type. This result suggests that the Fermi surface instability in $PrRu_4P_{12}$ is coupled with the multipole degrees of freedom of the $4f$ electrons in $PrRu_4P_{12}$, resulting in a metal-insulator transition preserving cubic symmetry. Such a multipole-based interpretation of the cubic-cubic transition is useful not only when multipoles are obviously involved in the electronic properties, as in the case of $PrRu_4P_{12}$, but also for understanding complex electronic phenomena and predicting the feasibility of nontrivial responses to external fields. For example, the cubic-cubic transition at $T_Q$ of $GaNb_4Se_8$ has been interpreted as a "molecular" quadrupole-ordered state of the $Nb_4Se_4$ clusters [128].

Thus, cubic materials that exhibit electronic phase transitions that preserve their cubic symmetry are expected to exhibit various unique electronic phenomena. However, significantly few materials exhibit such a phase transition, whose numbers are lesser than those of the electrically conducting pyrochlore oxides, spinel compounds with half-integer valence, and V and Ti compounds exhibiting triangular trimer formation, covered in Chapters 4-1, 4-2, and 4-3, respectively. The development of novel cubic-cubic transition materials is desired to discover outstanding quantum phenomena originating from cubic symmetry.

5. Summary

The research on the physical properties of transition metal oxides has spanned over several years. Multiple factors that play essential roles in the emergence of the superior physical properties have been identified. The β-pyrochlore oxide $CsW_2O_6$ exhibits a multitude of these factors. Among them, (i) geometrical frustration of the pyrochlore structure, (ii) moderately strong electron correlation and an incoherent metallic state, (iii) Jahn–Teller distortion and orbital order, (iv) charge order and metal-insulator transition, and (v) the formation of molecules in solids were discussed in this review. Other factors such as (vi) the possible formation of a complete flat band, (vii) rattling of Cs atoms and anharmonic phonons, and (viii) the topology of electronic states in Phase II may also play important roles; however, they are not covered in this review. Particularly, factor (vi) is a new mechanism that forms a flat band based on the strong spin–orbit coupling [138]. I hope that studies on $CsW_2O_6$ will expand, both theoretically and experimentally.

As a result of the interplay among these factors, the electronic properties that emerged at the phase transition between Phase I and Phase II at $T_t = 215$ K of $CsW_2O_6$ are unique from various perspectives, which are elaborated as follows: (a) an electronic phase transition preserving the cubic symmetry, (b) three-dimensional nesting of Fermi surfaces in the cubic system, (c) a charge order utilizing the fractional valence to avoid geometrical frustration, and (d) equilateral-triangular $W_3$ molecules composed of the 3c2e bonds. However, unsolved issues also persist in $CsW_2O_6$, such as the inconsistency between the experimental results related to the charge order, origin of the incoherency in Phase I, and crystal structure of Phase III. The synthesis of single-crystalline and sintered samples of $CsW_2O_6$, which are required for further studies, was described in Chapter 2.

Because $CsW_2O_6$ exhibits outstanding features from various perspectives, the range of related materials for comparison was vast. In Chapter 4, $CsW_2O_6$ was compared with as many materials as possible to provide a bird's-eye view of this topic, and the high potential of the materials discussed here for achieving interesting electronic phenomena has been reaffirmed. Further material exploration is expected to lead to the discovery of new materials with unique features beyond those of $CsW_2O_6$.

**Acknowledgments**

The author is grateful to K. Niki, R. Mitoka, Y. Yokoyama, K. Takenaka, H. Amano, T. Fujii, N. Katayama, H. Sawa, H. Harima, T. Hasegawa, N. Ogita, Y. Tanaka, M. Takigawa, K. Takehana, Y. Imanaka, Y. Nakamura, H. Kishida, R. Nakamura, D. Takegami, A. Medendez-Sans, L. H. Tjeng, M. Okawa, T. Miyashino, N. L. Saini, M. Kitamura, D. Shiga, H. Kumigashira, M. Yoshimura, K.-D. Tsuei, and T. Mizokawa for their helpful contributions on the studies on $CsW_2O_6$. The author is also grateful to Y. Yamakawa, A. Yamakage, Y. Motome, A. Koda, R. Kadono, J. Matsuno, D. Hirai, M. Shiomi, H. Nakai, C. Hotta, S. Kitou, R. Okuma, and K. Kojima for the helpful discussions. The author is grateful for the collaboration with Y. Nagao, J. Yamaura, M. Ichihara, Z. Hiroi, and M. Yoshida during the early stages of this study. This work was partially conducted under the Visiting Program of Institute for Solid State Physics, the University of Tokyo, and supported by JSPS KAKENHI (Grant Numbers: 16H03848, 18H04314, 19H05823, 23H01831, and 23K26524), JST ASPIRE (Grant Number: JPMJAP2314), the Research Foundation for the Electrotechnology of Chubu, and the Asahi Glass Foundation.

**References**

[1] H. R. Gaertner, Neues Jahrbuch für Mineralogie, Beilageband **61A**, 1 (1930).

[2] S. Yonezawa, Y. Muraoka, and Z. Hiroi, J. Phys. Soc. Jpn. **73**, 1655 (2004).

[3] M. A. Subramanian, G. Aravamudan, and G. V. Subba Rao, Prog. Solid St. Chem. **15**, 55 (1983).

[4] S. Yonezawa, Y. Muraoka, Y. Matsushita, and Z. Hiroi, J. Phys. Soc. Jpn. **73**, 819 (2004).

[5] W. Kurtz and S. Roth, Physica **86-88b**, 715 (1977).

[6] J. S. Gardner, M. J. P. Gingras, and J. E. Greedan, Rev. Mod. Phys. **82**, 53 (2010).

[7] M. J. Harris, S. T. Bramwell, D. F. McMorrow, T. Zeiske, and K. W. Godfrey, Phys. Rev. Lett. **79**, 2554 (1997).

[8] A. P. Ramirez, A. Hayashi, R. J. Cava, R. Siddharthan, and B. S. Shastry, Nature **399**, 333 (1999).

[9] J. Yamaura, K. Ohgushi, H. Ohsumi, T. Hasegawa, I. Yamauchi, K. Sugimoto, S. Takeshita, A. Tokuda, M. Takata, M. Udagawa, M. Takigawa, H. Harima, T. Arima, and Z. Hiroi, Phys. Rev. Lett. **108**, 247205 (2012).

[10] K. Tomiyasu, K. Matsuhira, K. Iwasa, M. Watahiki, S. Takagi, M. Wakeshima, Y. Hinatsu, M. Yokoyama, K. Ohoyama, and K. Yamada, J. Phys. Soc. Jpn. **81**, 034709 (2012).

[11] H. Sagayama, D. Uematsu, T. Arima, K. Sugimoto, J. J. Ishikawa, E. O'Farrell, and S. Nakatsuji, Phys. Rev. B **87**, 100403 (2013).

[12] Z. Hiroi, J. Yamaura, T. Hirose, I. Nagashima, and Y. Okamoto, APL Mater. **3**, 041501 (2015).

[13] A. W. Sleight, J. L. Gillson, J. F. Weiher, and W. Bindloss, Solid State Commun. **14**, 357 (1974).

[14] W. Klein, R. K. Kremer, and M. Jansen, J. Mater. Chem. **17**, 1356 (2007).

[15] A. Yamamoto, P. A. Sharma, Y. Okamoto, A. Nakao, H. Aruga-Katori, S. Niitaka, D. Hashizume, and H. Takagi, J. Phys. Soc. Jpn. **76**, 043703 (2007).

[16] K. Matsuhira, M. Wakeshima, R. Nakanishi, T. Yamada, A. Nakamura, W. Kawano, S. Takagi, Y. Hinatsu, J. Phys. Soc. Jpn. **76**, 043706 (2007).

[17] K. Matsuhira, M. Wakeshima, Y. Hinatsu, and S. Takagi, J. Phys. Soc. Jpn. **80**, 094701 (2011).

[18] A. W. Sleight and R. J. Bouchard, Solid State Chemistry, Proceedings of 5th Materials Research Symposium, July 1972, NBS Spec. Publ. 364, p. 227.

[19] T. Takeda, M. Nagata, H. Kobayashi, R. Kanno, Y. Kawamoto, M. Takano, T. Kamiyama, F. Izumi, and A. W. Sleight, J. Solid State Chem. **140**, 182 (1998).

[20] T. Takeda, R. Kanno, Y. Kawamoto, M. Takano, F. Izumi, A. W. Sleight, and A. W. Hewat, J. Mater. Chem. **9**, 215 (1999).

[21] M. Hanawa, Y. Muraoka, T. Tayama, T. Sakakibara, J. Yamaura, and Z. Hiroi, Phys. Rev. Lett. **87**, 187001 (2001).

[22] H. Sakai, K. Yoshimura, H. Ohno, H. Kato, S. Kambe, R. E. Walstedt, T. D. Matsuda, Y. Haga, and Y. Onuki, J. Phys.: Condens. Matter **13**, L785 (2001).

[23] S. Yonezawa, Y. Muraoka, Y. Matsushita, and Z. Hiroi, J. Phys.: Condens. Matter **16**, L9 (2004).

[24] X. Wan, A. M. Turner, A. Vishwanath, and S. Y. Savrasov, Phys. Rev. B **83**, 205101 (2011).

[25] K. Y. Yang, Y. M. Lu, and Y. Ran, Phys. Rev. B **84**, 075129 (2011).

[26] R. J. Cava, R. S. Roth, T. Siegrist, B. Hessen, J. J. Krajewski, and W. F. Peck, Jr., J. Solid State Chem. **103**, 359 (1993).

[27] D. Hirai, M. Bremholm, J. M. Allred, J. Krizan, L. M. Schoop, Q. Huang, J. Tao, and R. J. Cava, Phys. Rev. Lett. **110**, 166402 (2013).

[28] S. V. Streltsov, I. I. Mazin, R. Heid, and K. P. Bohnen, Phys. Rev. B **94**, 241101 (2016).

[29] T. Soma, K. Yoshimatsu, K. Horiba, H. Kumigashira, and A. Ohtomo, Phys. Rev. Mater. **2**, 115003 (2018).

[30] Y. Okamoto, H. Amano, N. Katayama, H. Sawa, K. Niki, R. Mitoka, H. Harima, T. Hasegawa, N. Ogita, Y. Tanaka, M. Takigawa, Y. Yokoyama, K. Takehana, Y. Imanaka, Y. Nakamura, H. Kishida, and K. Takenaka, Nat. Commun. **11**, 3144 (2020).

[31] Y. Okamoto, K. Niki, R. Mitoka, and K. Takenaka, J. Phys. Soc. Jpn. **89**, 124710 (2020).

[32] R. Nakamura, D. Takegami, A. Melendez-Sans, L. H. Tjeng, M. Okawa, T. Miyashino, N. L. Saini, M. Kitamura, D. Shiga, H. Kumigashira, M. Yoshimura, K.-D. Tsuei, Y. Okamoto, and T. Mizokawa, Phys. Rev. B **106**, 195104 (2022).

[33] S. Uchida, T. Ido, H. Takagi, T. Arima, Y. Tokura, and S. Tajima, Phys. Rev. B **43**, 7942 (1991).

[34] M. Imada, A. Fujimori, and Y. Tokura, Rev. Mod. Phys. **70**, 1039 (1998).

[35] D. N. Basov, R. D. Averitt, D. van der Marel, M. Dressel, and K. Haule, Rev. Mod. Phys. **83**, 471 (2011).

[36] M. Shiomi, K. Kojima, N. Katayama, Y. Okamoto, and H. Sawa, JPS 2020 Autumn Meeting (2020).

[37] Y. Okamoto, M. Nohara, H. Aruga-Katori, and H. Takagi, Phys. Rev. Lett. **99**, 137207 (2007).

[38] Y. Okamoto, G. J. Nilsen, J. P. Attfield, and Z. Hiori, Phys. Rev. Lett. **110**, 097203 (2013).

[39] I. D. Brown and D. Altermatt, Acta Crystallogr. **B41**, 244–247 (1985).

[40] Y. Okamoto, M. Mori, N. Katayama, A. Miyake, M. Tokunaga, A. Matsuo, K. Kindo, and K. Takenaka, J. Phys. Soc. Jpn. **87**, 034709 (2018).

[41] G. J. Thorogood, P. J. Saines, B. J. Kennedy, R. L. Withers, and M. M. Elcombe, Mater. Res. Bull. **43**, 787–795 (2008).

[42] M. Tachibana, Y. Kohama, T. Shimoyama, A. Harada, T. Taniyama, M. Itoh, H. Kawaji, and T. Atake, Phys. Rev. B **73**, 193107 (2006).

[43] H. Sakai, H. Ohno, N. Oba, M. Kato, and K. Yoshimura, Physica B **329-333**, 1038 (2003).

[44] Y. Hirata, M. Nakajima, Y. Nomura, H. Tajima, Y. Matsushita, K. Asoh, Y. Kiuchi, A. G. Eguiluz, R. Arita, T. Suemoto, and K. Ohgushi, Phys. Rev. Lett. **110**, 187402 (2013).

[45] R. Jin, J. He, J. R. Thompson, M. F. Chisholm, B. S. Sales, and D. Mandrus, J. Phys.: Condens. Matter **14**, L117 (2002).

[46] M. Hanawa, J. Yamaura, Y. Muraoka, and F. Sakai, and Z. Hiroi, J. Phys. Chem. Solids **63**, 1027 (2002).

[47] Z. Hiroi, S. Yonezawa, J. Yamaura, T. Muramatsu, and Y. Muraoka, J. Phys. Soc. Jpn. **74**, 1682 (2005).

[48] J. Yamaura, M. Takigawa, O. Yamamuro, and Z. Hiroi, J. Phys. Soc. Jpn. **79**, ,043601 (2010).

[49] K. Sasai, M. Kofu, R. M. Ibberson, K. Hirota, J. Yamaura, Z. Hiroi, and O. Yamamuro, J. Phys.: Condens. Matter **22**, 015403 (2009).

[50] Z. Hiroi, J. Yamaura, and K. Hattori, J. Phys. Soc. Jpn. **81**, 011012 (2012).

[51] Y. Shimakawa, Y. Kubo, and T. Manako, Nature **379**, 53 (1996).

[52] M. A. Subramanian, B. H. Toby, A. P. Ramirez, W. J. Marshall, A. W. Sleight, and G. H. Kwei, Science **273**, 81 (1996).

[53] Y. Taguchi and Y. Tokura, Phys. Rev. B **60**, 10280 (1999).

[54] S. Yoshii, S. Iikubo, T. Kageyama, K. Oda, Y. Kondo, K. Murata, and M. Sato, J. Phys. Soc. Jpn. **69**, 3777 (2000).

[55] J. Reading, S. Gordeev, and M. T. Weller, J. Mater. Chem. **12**, 646 (2002).

[56] M. Tachibana, H. Kawaji, and T. Atake, Solid State Commun. **131**, 745 (2004).

[57] K. Kataoka, D. Hirai, A. Koda, R. Kadono, T. Honda, and Z. Hiroi, J. Phys.: Condens. Matter **34**, 135602 (2022).

[58] R. Wang and A. W. Sleight, Mat. Res. Bull. **33**, 1005 (1998).

[59] Y. Y. Jiao, J. P. Sun, P. Shahi, Q. Cui, X. H. Yu, Y. Uwatoko, B. S. Wang, J. A. Alonso, H. M. Weng, and J.-G. Cheng, Phys. Rev. B **98**, 075118 (2018).

[60] T. Munenaka and H. Sato, J. Phys. Soc. Jpn. **75**, 103801 (2006).

[61] Y. Nakayama, Y. Okamoto, D. Hirai, and K. Takenaka, J. Phys. Soc. Jpn. **91**, 125002 (2022).

[62] M. Miyazaki, R. Kadono, K. H. Satoh, M. Hiraishi, S. Takeshita, A. Koda, A. Yamamoto, and H. Takagi, Phys. Rev. B **82**, 094413 (2010).

[63] M. Yoshida, M. Takigawa, A. Yamamoto, and H. Takagi, J. Phys. Soc. Jpn. **80**, 034705 (2011).

[64] S. M. Disseler, Chetan Dhital, T. C. Hogan, A. Amato, S. R. Giblin, Clarina de la Cruz, A. Daoud-Aladine, Stephen D. Wilson, and M. J. Graf, Phys. Rev. B **85**, 174441 (2012).

[65] J. van Duijn, R. Ruiz-Bustos, and A. Daoud-Aladine, Phys. Rev. B **86**, 214111 (2012).

[66] J. E. Greedan, M. Sato, N. Ali, and W. R. Datars, J. Solid State Chem. **68**, 300 (1987).

[67] P. Hill, S. Labroo, X. Zhang, and N. Ali, J. Less-Common Metals **149**, 327 (1989).

[68] P. C. Donohue, J. M. Longo, R. D. Rosenstein, and L. Katz, Inorg. Chem. **4**, 1152 (1965).

[69] C. Michioka, Y. Kataoka, H. Ohta, and K. Yoshimura, J. Phys.: Condens. Matter **23**, 445602 (2011).

[70] N. P. Raju, J. E. Greedan, and M. A. Subramanian, Phys. Rev. B **49**, 1086 (1994).

[71] J. Dai, Y. Yin, X. Wang, X. Shen, Z. Liu, X. Ye, J. Cheng, C. Jin, G. Zhou, Z. Hu, S. Weng, X. Wan, and Y. Long, Phys. Rev. B **97**, 085103 (2018).

[72] S. Nakatsuji, Y. Machida, Y. Maeno, T. Tayama, T. Sakakibara, J. van Duijn, L. Balicas, J. N. Millican, R. T. Macaluso, and J. Y. Chan, Phys. Rev. Lett. **96**, 087204 (2006).

[73] H. L. Feng, C.-J. Kang, Z. Deng, M. Croft, S. Liu, T. A. Tyson, S. H. Lapidus, C. E. Frank, Y. Shi, C. Jin, D. Walker, G. Kotliar, and M. Greenblatt, Inorg. Chem. **60**, 4424 (2021).

[74] S. Lee, J.-G. Park, D. T. Adroja, D. Khomskii, S. Streltsov, K. A. McEwen, H. Sakai, K. Yoshimura, V. I. Anisimov, D. Mori, R. Kanno, and R. Ibberson, Nat. Mater. **5**, 471 (2006).

[75] Z. Hiroi, J. Yamaura, T. C. Kobayashi, Y. Matsubayashi, and D. Hirai, J. Phys. Soc. Jpn. **87**, 024702 (2018).

[76] Y. Matsubayashi, K. Sugii, D. Hirai, Z. Hiroi, T. Hasegawa, S. Sugiura, H. T. Hirose, T. Terashima, and S. Uji, Phys. Rev. B **101**, 205133 (2020).

[77] S. Uji, S. Sugiura, H. T. Hirose, T. Terashima, Y. Matsubayashi, D. Hirai, Z. Hiroi, and T. Hasegawa, Phys. Rev. B **102**, 155131 (2020).

[78] H. T. Hirose, T. Terashima, D. Hirai, Y. Matsubayashi, N. Kikugawa, D. Graf, K. Sugii, S. Sugiura, Z. Hiroi, and Shinya Uji, Phys. Rev. B **105**, 035116 (2022).

[79] J. S. Lee, Y. S. Lee, K. W. Kim, T. W. Noh, J. Yu, T. Takeda, and R. Kanno, Phys. Rev. B **64**, 165108 (2001).

[80] M. Kasuya, Y. Tokura, T. Arima, H. Eisaki, and S. Uchida, Phys. Rev. B **47**, 6197 (1993).

[81] Y. Taguchi, Y. Tokura, T. Arima, and F. Inaba, Phys. Rev. B **48**, 511 (1993).

[82] K. Ueda, J. Fujioka, Y. Takahashi, T. Suzuki, S. Ishiwata, Y. Taguchi, and Y. Tokura, Phys. Rev. Lett. **109**, 136402 (2012).

[83] C. H. Sohn, Hogyun Jeong, Hosub Jin, Soyeon Kim, L. J. Sandilands, H. J. Park, K. W. Kim, S. J. Moon, Deok-Yong Cho, J. Yamaura, Z. Hiroi, and T. W. Noh, Phys. Rev. Lett. **115**, 266402 (2015).

[84] E. J. W. Verwey, Nature **144**, 327 (1939).

[85] N. Tsuda, K. Nasu, A. Fujimori, and K. Siratori, Electronic Conduction in Oxides (pp 243–270, Springer, 2000).

[86] E. J. W. Verwey and P. W. Haaymann, Physica **8**, 979 (1941).

[87] E. J. W. Verwey, P. W. Haaymann, and F. C. Romeijn, J. Chem. Phys. **15**, 174 (1947).

[88] P. W. Anderson, Phys. Rev. **102**, 1008 (1956).

[89] J. P. Wright, J. P. Attfield, and P. G. Radaelli, Phys. Rev. Lett. **87**, 266401 (2001).

[90] P. G. Radaelli, Y. Horibe, M. J. Gutmann, H. Ishibashi, C. H. Chen, R. M. Ibberson, Y. Koyama, Y. S. Hor, V. Kiryukhin, S. W. Cheong, Nature **416**, 155 (2002).

[91] T. Furubayashi, T. Matsumoto, T. Hagino, and S. Nagata, J. Phys. Soc. Jpn. **63**, 3333 (1994).

[92] Y. Horibe, M. Shingu, K. Kurushima, H. Ishibashi, N. Ikeda, K. Kato, Y. Motome, N. Furukawa, S. Mori, and T. Katsufuji, Phys. Rev. Lett. **96**, 086406 (2006).

[93] K. Matsuno, T. Katsufuji, S. Mori, Y. Moritomo, A. Machida, E. Nishibori, M. Takata, M. Sakata, N. Yamamoto, and H. Takagi, J. Phys. Soc. Jpn. **70**, 1456 (2001).

[94] Z. Hiroi, Z, Prog. Solid State Chem. 43, 47–69 (2015).

[95] J. P. Attfield, APL Mater. **3**, 041510 (2015).

[96] M. S. Senn, J. P. Wright, and J. P. Attfield, Nature **481**, 173–176 (2012).

[97] K. Takubo, S. Hirata, J.-Y. Son, J. W. Quilty, T. Mizokawa, N. Matsumoto, and S. Nagata, Phys. Rev. Lett. **95**, 246401 (2005).

[98] D. I. Khomskii and T. Mizokawa, Phys. Rev. Lett. **94**, 156402 (2005).

[99] A. J. Browne, S. A. J. Kimber, and J. P. Attfield, Phys. Rev. Mater. **1**, 052003 (2017).

[100] M. Matsuda, H. Ueda, A. Kikkawa, Y. Tanaka, K. Katsumata, Y. Narumi, T. Inami, Y. Ueda, S.-H. Lee, Nat. Phys. **3**, 397 (2007).

[101] H. Kawai, M. Tabuchi, M. Nagata, H. Tukamoto, and A. R. West, J. Mater. Chem. **8**, 1273 (1998).

[102] P. Strobel, A. I. Palos, M. Anne, and F. L. Cras, J. Mater. Chem. **10**, 429 (2000).

[103] P. Badica, T. Kondo, and Kazumasa Togano, J. Phys. Soc. Jpn. **74**, 1014 (2005).

[104] M. Shiomi, K. Kojima, N. Katayama, S. Maeda, J. A. Schneeloch, S. Yamamoto, K. Sugimoto, Y. Ohta, D. Louca, Y. Okamoto, and H. Sawa, Phys. Rev. B **105**, L041103 (2022).

[105] Y. Okamoto, S. Niitaka, M. Uchida, T. Waki, M. Takigawa, Y. Nakatsu, A. Sekiyama, S. Suga, R. Arita, and H. Takagi, Phys. Rev. Lett. **78**, 033701 (2009).

[106] M. Taguchi, A. Chainani, S. Ueda, M. Matsunami, Y. Ishida, R. Eguchi, S. Tsuda, Y. Takata, M. Yabashi, K. Tamasaku, Y. Nishino, T. Ishikawa, H. Daimon, S. Todo, H. Tanaka, M. Oura, Y. Senba, H. Ohashi, and S. Shin, Phys. Rev. Lett. **115**, 256405 (2015).

[107] Y. Nakatsu, A. Sekiyama, S. Imada, Y. Okamoto, S. Niitaka, H. Takagi, A. Higashiya, M. Yabashi, K. Tamasaku, T. Ishikawa, and S. Suga, Phys. Rev. B **83**, 115120 (2011).

[108] T. A. Hewston and B. L. Chamberland, J. Solid State Chem. **59**, 168 (1985).

[109] T. A. Hewston and B. L. Chamberland, J. Solid State Chem. **65**, 100 (1986)

[110] M. Onoda, T. Naka, and H. Nagasawa, J. Phys. Soc. Jpn. **60**, 2550 (1991).

[111] J. B. Goodenough, G. Dutta, and A. Manthiram, Phys. Rev. B **43**, 10170 (1991).

[112] J. B. Goodenough, Magnetism and the Chemical Bond (pp 266–270, John Wiley & Sons, 1963).

[113] H. F. Pen, J. van den Brink, D. I. Khomskii, and G. A. Sawatzky, Phys. Rev. Lett. **78**, 1323 (1997).

[114] N. Katayama, M. Uchida, D. Hashizume, S. Niitaka, J. Matsuno, D. Matsumura, Y. Nishihata, J. Mizuki, N. Takeshita, A. Gauzzi, M. Nohara, and H. Takagi, Phys. Rev. Lett. **103**,

146405 (2009).

[115] N. Katayama, K. Kojima, T. Yamaguchi, S. Hattori, S. Tamura, K. Ohara, S. Kobayashi, K. Sugimoto, Y. Ohta, K. Saitoh, and H. Sawa, npj Quantum Mater. **6**, 16 (2021).

[116] D. J. Hinz, G. Meyer, T. Dedecke, and W. Urland. Angew. Chem. Int. Ed. Engl. **34**, 71 (1995).

[117] N. Hänni, M. Frontzek, J. Hauser, D. Cheptiakov, and K. Krämer, Z. Anorg. Allg. Chem. **643**, 2063 (2017).

[118] Z. A. Kelly, T. T. Tran, and T. M. McQueen, Inorg. Chem. **58**, 11941 (2019).

[119] T. Kajita, Y. Obata, Y. Kakesu, Y. Imai, Y. Shmizu, M. Itoh, H. Kuwahara, and T. Katsufuji, Phys. Rev. B **96**, 245126 (2017).

[120] J. Miyazaki, K. Matsudaira, Y. Shimizu, M. Itoh, Y. Nagamine, S. Mori, J. E. Kim, K. Kato, M. Takata, and T. Katsufuji, Phys. Rev. Lett. **104**, 207201 (2010).

[121] T. Kajita, H. Kuwahara, S. Mori, and T. Katsufuji, Phys. Rev. Research **3**, 033046 (2021).

[122] U. Müller, Inorganic Structural Chemistry (pp. 128–149, John Wiley & Sons, Chichester, UK, 2007).

[123] D. Hirai, K. Kojima, N. Katayama, M. Kawamura, D. Nishio-Hamane, and Zenji Hiroi, J. Am. Chem. Soc. **144**, 17857 (2022).

[124] T. Oka, Phys. Rev. Lett. **45**, 531 (1980).

[125] A. R. West, Solid State Chemistry and Its Applications (pp. 404–411, Wiley, John Wiley & Sons, Chichester, UK, 2022)

[126] U. Müller, Inorganic Structural Chemistry (pp. 157–165, John Wiley & Sons, Chichester, UK, 2007).

[127] W. I. F. David, R. M. Ibberson, J. C. Matthewman, K. Prassides, T. J. S. Dennis, J. P. Hare, H. W. Kroto, R. Taylor, and D. R. M. Walton, Nature **353**, 147 (1991).

[128] H. Ishikawa, T. yajima, A. Matsuo, Y. Ihara, and K. Kindo, Phys. Rev. Lett. **124**, 227202 (2020).

[129] C. Sekine, T. Uchiumi, I. Shirotani, and T. Yagi, Phys. Rev. Lett. **79**, 3218 (1997).

[130] H.-S. Kim, J. Im, M. J. Han, and H. Jin, Nat. Commun, **5**, 3988 (2014).

[131] E. Ruff, S. Widmann, P. Lunkenheimer, V. Tsurkan, S. Bordács, I. Kézsmárki, A. Loidl, Sci. Adv. **1**, e1500916 (2015).

[132] Y. Fujima, N. Abe, Y. Tokunaga, and T. Arima, Phys. Rev. B **95**, 180410 (2017).

[133] S. Kitou, M. Gen, Y. Nakamura, Y. Tokunaga, and T. Arima, Chem. Mater. **36**, 2993 (2024).

[134] C. H. Lee, H. Matsuhata, A. Yamamoto, T. Ohta, H. Takazawa1, K. Ueno, C. Sekine, I. Shirotani, and T. Hirayama, J. Phys.: Condens. Matter **13**, L45 (2001).

[135] H. Harima, K. Takegahara, S. H. Curnoe, and K. Ueda, J. Phys. Soc. Jpn. **71**, 70 (2002).

[136] C. H. Lee, H. Matsuhata, H. Yamaguchi, C. Sekine, K. Kihou, T. Suzuki, T. Noro, and I. Shirotani, Phys. Rev. B **70**, 153105 (2004).

[137] I. Shirotani, T. Adachi, K. Tachi, S. Todo, K. Nozawa, T. Yagi, and M. Kinoshita, J. Phys. Chem. Solids **57**, 211 (1996).

[138] H. Nakai and C. Hotta, Nat. Commun. **13**, 579 (2022).

*email: yokamoto@issp.u-tokyo.ac.jp